\documentstyle[12pt,epsfig]{article}
\newcommand{\beq}{\begin{equation}}
\newcommand{\eeq}{\end{equation}}
\newcommand{\beqn}{\begin{eqnarray}}
\newcommand{\eeqn}{\end{eqnarray}}
\newcommand{\stackM}{\stackrel{\scriptstyle >}{{ }_{\sim}}}
\newcommand{\stackm}{\stackrel{\scriptstyle <}{{ }_{\sim}}} 
\begin{document}
\thispagestyle{empty}
\def\pubnum{401}
\def\data{February, 1997}

\def\UUAABB{\hfill\hbox{
    \vrule height0pt width2.5in
    \vbox{\hbox{\rm 
     UAB-FT-\pubnum
    }\break\hbox{\data\hfill}
    \break\hbox{hep-ph/9702390\hfill} 
    \hrule height2.7cm width0pt}
   }}   
\UUAABB
\vspace{3cm}
\begin{center}
\begin{large}
\begin{bf}
TOP AND HIGGS DECAYS: LOOKING FOR QUANTUM SUSY SIGNATURES
IN HADRON COLLIDERS
\footnote{Updated analysis based on the talks
presented at SUSY 96, Univ. of Maryland,
College Park, USA, May 29th-June 1st 1996; and at CRAD 96,
Cracow, Poland, 1-5 August 1996.}\\
\end{bf}
\end{large}
\vspace{0.75cm}
Joan SOL{\`A}\\
\vspace{0.2cm} 
Grup de F{\'\i}sica Te{\`o}rica\\ 
and\\ 
Institut de F{\'\i} sica d'Altes Energies\\ 
\vspace{0.2cm} 
Universitat Aut{\`o}noma de Barcelona\\
08193 Bellaterra (Barcelona), Catalonia, Spain\\
\end{center}
\vspace{0.25cm}
\hyphenation{super-symme-tric re-la-ti-ve-ly par-ti-ci-pa-ting}
\hyphenation{com-pe-ti-ti-ve re-nor-ma-li-za-tion}
\hyphenation{con-si-de-ra-tions}
\begin{center}

{\bf ABSTRACT}
\end{center}
\begin{quotation}
\noindent
I present an updated discussion of the one-loop 
corrections to the standard top quark decay, 
$t\rightarrow W^{+}\, b$, and to the charged Higgs mode, 
$t\rightarrow H^{+}\, b$, within the framework of the MSSM. 
These higher order contributions to the top width are compared with 
the direct top decays into $2$ and $3$ real SUSY particles.
It turns out
that whereas the SUSY effects on $t\rightarrow W^{+}\, b$ add up to
those from standard QCD, they tend most likely to 
erase the conventional QCD corrections to $t\rightarrow H^{+}\, b$.
However, in the latter case the full (strong plus electroweak)
SUSY quantum effects are generally so
large that the QCD corrections effectively 
come out with the ``wrong sign''. A qualitatively distinct, and
quantitatively significant, ``quantum signature'' of that type 
might constitute a most sought-after imprint of ``virtual''
Supersymmetry, which remains sizeable enough even in the absence
of any direct top decay into SUSY particles. Similar considerations 
are made for the hadronic partial widths of the Higgs bosons of the MSSM.
While a first valuable test of these effects could
possibly be performed at the upgraded Tevatron, a more precise verification
would most probably be carried out in future experiments at the LHC.  
\end{quotation}

\baselineskip=6.5mm  
 
\newpage
\vspace{0.75cm} 
\begin{Large}
{\bf 1. Introduction}
\end{Large}
\vspace{0.5cm}

For a long time, Supersymmetry (SUSY)\,\cite{WessZumino} has been an 
extremely tantalizing candidate to extend the quantum field
theoretical structure of the Standard Model (SM)
of the strong and electroweak interactions while keeping all the
necessary ingredients insuring internal consistency, such as
gauge invariance and renormalizability. A major goal of SUSY is to
produce a unified theory of all the interactions, including gravity.
However, this theory calls for the existence of a host of new particles,
the so-called supersymmetric particles (``sparticles''). They must be found
to substantiate the theory. 
At present, the simplest and most popular realization of this idea,
namely the Minimal Supersymmetric Standard Model (MSSM)\,\cite{MSSM},
is being thoroughly scrutinized by experiment and it has successfully
passed all the tests up to now. As a matter of fact, 
the global fit analyses to a huge number of indirect precision
data within the MSSM are slightly worse than in the SM, due to the large
number of parameters involved, but they are still comparable
to those in the SM\,\cite{WdeBoer},
a situation which is certainly {\it not} shared by 
all known alternative extensions of
the SM, like any of the multifarious extended Technicolour models.
 
In the near and middle future, with the upgrade of the
Tevatron, the advent of the LHC, and the possible
construction of an $e^+\,e^-$ supercollider (NLC), 
new results on top quark physics, and perhaps also on Higgs
physics, will be obtained in interplay
with Supersymmetry that may
be extremely helpful to complement the precious information
already collected at LEP from $Z$-boson physics\,\cite{LEPEWWG}.  

In this talk I propose to dwell on the phenomenology of
top quark and Higgs boson
decays with an eye on these future developments. 
A recent reanalysis of experimental studies on Higgs
detection possibilities at the LHC has shown that
 the design of the LHC detectors 
is also very suitable to tag signals associated to production of
supersymmetric particles\,\cite{Fernandez}.  
However, even in the absence of direct detection of SUSY particles,
there remain important SUSY quantum effects that have not been
fully explored and that could
play a fundamental role in the search for new physics.
Here I wish to elaborate mainly on the
kind of ``high precision'' measurements
that may be performed on top quark
and Higgs boson observables in hadron colliders in order to catalyze
a potential discovery of ``virtual'' Supersymmetry.

But, why hadron colliders?...Well, as I have said, in part because
the Tevatron has been doing a very good job, and currently the
machine is being significantly improved (energy and luminosity) for the 
next runs (Run II, TeV$33$ Project,...)\,\cite{Willenbrock}, and 
also because of the superb LHC program that we have ahead\,\cite{Atlas}.
Now, however important these
reasons may be, there are also additional, more theoretical, 
arguments beyond the actual machines' programmes and upgrading
that should be taken into account. In the following
I will expand a bit on these additional features.

By ``high precision'' measurements I mean, as usual, measurements
sensitive to quantum effects.   
Unfortunately (but, of course, naturally) quantum effects are
generally tiny since they are computed from perturbation theory.
And, due to this fact, it is not easy to clutch at them unless one
chooses a very clean and smart experimental environment, typically an
$e^+\,e^-$ machine like LEP or SLC. Machines like these did offer us
in the past (and still partly at present)
an immaculate laboratory for electroweak physics where 
high precision experiments can be performed. This is both
because of the relative simplicity of the $e^+\,e^-$ interactions and because
of the possibility to collect
a lot of event statistics. Not too surprisingly, therefore,
LEP and SLC running at the $Z$-pole mass
have revealed themselves as the greatest possible scenarios yet 
imagined for high precision experiments ever. Regrettably, this magnificent
setup has a serious drawback:
it can work at full rate only at a fixed energy $\sqrt{s}=M_Z$.
And, to our distress, after all the scans searching for
traces of new physics around that 
energy  have been accomplished, they have absolutely failed to unravel 
anything new at all. In this respect the situation has 
not improved any better\,\cite{RMiquel} neither at LEP 1.61 nor at
LEP 1.72 thus far, where in spite of the higher energy
and high performance of the machine
the event statistics became, of course, substantially diminished.

Admittedly, it would suffice collecting a few ``real'' non-SM events
to ``prove'', or at
least to reasonably contend, that new physics may be around. And in
this sense, ``direct'' new physics at LEP/SLC would be very fine. In the absence of
this possibility up to the present moment, one would naturally rush to quest
for ``quantum signs'' of new physics, in particular of Supersymmetry, by means
of the indirect method of high precision measurements. 
Notwithstanding, this alternative 
approach to searching for Supersymmetry in these machines,
apparently does not lead to too promising expectancies either.
And this will not change during the LEP/SLC lifetime. 
In the end a more fundamental
fact must be undermining our experimental ability to search for 
``virtual'' Supersymmetry in an
$e^+\,e^-$ machine. 
It must be that virtual SUSY effects are doomed to be small in an $e^+\,e^-$
environment. Of course,
in the future, NLC might resolve this embarrassment
very expeditiously on the basis of 
``brute force'' production of real sparticles 
whatever they mass
be in the range $\stackm 1\,TeV$. But in the meanwhile the method of
``quantum signatures'' should be seriously considered, already at the
next Tevatron run.  

It is a general fact that all sorts of SUSY effects 
fade away when we arbitrarily increase the sparticle masses. This 
is nothing else but an unavoidable consequence of the 
decoupling theorem (DT)\,\cite{DT}
applied to the MSSM. This theorem has been
well tested in the MSSM after a long experience of computations of 
supersymmetric quantum corrections to gauge boson observables,
like e.g. the $W$-mass\,\cite{GrifSol}-\cite{MW}, the $Z$-width\,\cite{GJSZ}
and all kinds of related observables, like $\sin^2\theta_W$,
 $R_b$, $R_c$, the 
various asymmetries etc\,\cite{Rb,WdeBoer}. 

That the DT holds in the MSSM is bound to reason; after all,
the source of SUSY breaking 
is independent -- or at least it is not directly related -- to the 
source of spontaneous symmetry
breaking (SSB) of the gauge theory. Indeed, 
the soft SUSY-breaking masses, which we denote collectively by $M_{SUSY}$,
are $SU(2)_L\times U(1)_Y$-invariant. Therefore, since
$M_{SUSY}$ can be arbitrarily large,
it can quickly dominate the physical masses of all the SUSY partners.  
In these asymptotic conditions, the sparticle masses
just increase because we are
increasing the size of a {\it dimensionful} parameter; and this is precisely
a sufficient condition for the decoupling theorem
to hold\,\cite{DT}.

Coming back to $e^+\,e^-$ phenomenology, experience shows us that
all kinds of SUSY quantum effects inherently bound to the physics of the
gauge bosons come out to be very small (namely, at the level of a few percent
or below); and what is more, they drop off very fast on increasing 
the characteristic low-energy electroweak
scale, $M_{SUSY}$, which determines the 
soft SUSY-breaking masses of the supersymmetric
partners. 
Therefore, aside from direct ``real'' sparticle production,
which is always an exciting possibility (in fact, the best conceivable SUSY
signature), there is little hope for
being able in the future to grasp a hint of SUSY from 
 ``virtual''signatures lying behind the
dynamics of gauge boson observables. 
We need to turn our attention to
a ``new'' scenario, namely, one in which ``quantum SUSY signatures'' can be 
significantly larger and effectively less subdued by the decoupling theorem.

Fortunately, this scenario is at our disposal: it is the physics of 
{\it real} top/bottom quarks in interplay with
Higgs bosons in hadron colliders (see below).
Of course, I am not suggesting that the
SUSY quantum effects generated in this sector of the theory may escape the 
consequences of the DT. I only  dare to say that there are smart 
situations where
those quantum effects are much larger (for the same values of the
sparticle masses), and that the decoupling rate  
is slower, than in the weak gauge boson case. 

\vspace{0.75cm} 
\begin{Large}
{\bf 2. $t\rightarrow W^+\,b$ versus $t\rightarrow H^+\,b$ in the MSSM
and the renormalization of $\tan\beta$}
\end{Large}
\vspace{0.5cm}

Why do we expect potentially large virtual SUSY signatures in top quark physics?.
It is well known that an exception to the DT is the phenomenon
of SSB, which the MSSM has in common with the SM.
Here, for fixed $M_{SUSY}$, we let dimensionless parameters (such as the
Yukawa couplings) go to infinity and the corresponding quantum corrections
do {\it not} decouple.
Now, in the MSSM, the spectrum of Higgs-like particles  and of
Yukawa couplings is far and away richer than in the SM. 
A crucial fact affecting the results of our work is that in such a framework the
bottom-quark Yukawa coupling may counterbalance the smallness of the bottom
mass, $m_b\simeq 4-5\,GeV$,
at the expense of a large value of $\tan\beta$ -- the ratio of the vacuum
expectation values (VEV's) of the two Higgs doublets -- the upshot being that
the top-quark and bottom-quark Yukawa couplings
(normalized with respect to the $SU(2)$ gauge coupling)
as they stand in the superpotential\,\cite{MSSM}, read
\begin{equation}
\lambda_t\equiv {h_t\over g}={m_t\over \sqrt{2}\,M_W\,\sin{\beta}}\;\;\;\;\;,
\;\;\;\;\; \lambda_b\equiv {h_b\over g}={m_b\over \sqrt{2}
\,M_W\,\cos{\beta}}\,,
\label{eq:Yukawas} 
\end{equation}
and can be of the same order of magnitude, perhaps even showing up 
in ``inverse hierarchy'': $h_t<h_b$ for $\tan\beta> m_t/m_b$. 
Clearly, both at large and small values of $\tan\beta$ the Yukawa couplings
(\ref{eq:Yukawas}) can be greatly enhanced as compared to the SM. 
For definiteness, in our numerical analysis we will use the (approximate) 
perturbatively safe range
\beq
\left({g\,m_t\over 2\,M_W}\simeq 0.7\right)\stackm\tan\beta
\stackm \left({2\,M_W\over g\,m_b}\simeq 60\right)\,,
\label{eq:tanbeta1}
\eeq
which can be fixed e.g. by tracking down the size of the squared
Yukawa couplings of the CP-odd Higgs boson,
$A^0$, to top and bottom quarks:
\beqn
& &\left({g\,m_b\,\tan\beta\over 2\,M_W}\right)^2
\stackm 1\ \ \ (\tan\beta\stackm 60)\,,\nonumber\\
& &\left({g\,m_t\,\cot\beta\over 2\,M_W}\right)^2
\stackm 1\ \ \ (\tan\beta\stackM 0.7)\,.
\label{eq:perturb}
\eeqn
Notice that the Yukawa couplings of $A^0$ are most suitable to fix the
allowed range of $\tan\beta$ since they do not
depend on the additional mixing angle, $\alpha$, as it would be the case for
the CP-even Higgs bosons $h^0,H^0$\,\cite{Hunter}. 

From the previous considerations, it should be clear that a wealth of
interesting new physics could potentially be unearthed from the study of
Higgs-quark-quark vertices in the MSSM,
basically (though not exclusively) those involving top and bottom quarks 
in combination with charged and neutral Higgs bosons. 
To start with, I shall review and update our work
\footnote{See the original references
\cite{GJSH}-\cite{CGGJS} and the proceedings' articles\,\cite{Proceedings}.} 
on one-loop quantum corrections to the top quark decays $t\rightarrow W^+\,b$ 
and $t\rightarrow H^+\,b$
mediated by the plethora of
supersymmetric partners, such as squarks, sleptons, gluinos,
chargino-neutralinos
and the various Higgs bosons of the MSSM,
and shall compare them with the 
standard QCD corrections.

Whereas a simple tree-level study of $t\rightarrow H^+\,b$
is insensitive to
the nature of the Higgs sector to which $H^+$ belongs,  
a careful study of the leading quantum
effects on that decay
could be the clue to unravel the potential
supersymmetric nature of the charged Higgs.
In particular, it should be useful
to distinguish it from a charged Higgs belonging to
a general two-Higgs-doublet model ($2$HDM). Now,
of course, part of these quantum effects (viz. the conventional QCD corrections) 
cannot distinguish the
structure of the underlying Higgs
model. Nevertheless, their knowledge is indispensable
to probe the existence of
additional sources of strong virtual corrections beyond the SM.

To evaluate the relevant quantum corrections,
we shall adopt the on-shell renormalization scheme\,\cite{BSH} where the
fine structure constant,
$\alpha$, and the masses of the gauge bosons, fermions and scalars are
the renormalized parameters ($\alpha$-scheme). 
Apart from the well-known  $t\,b\,W^{+}$ interaction,
the Lagrangian describing the $t\,b\,H^{+}$ vertex 
in the MSSM reads as follows: 
\beq
{\cal L}_{Htb}={g\,V_{tb}\over\sqrt{2}M_W}\,H^+\,\bar{t}\,
[m_t\cot\beta\,P_L + m_b\tan\beta\,P_R]\,b+{\rm h.c.}\,,
\label{eq:LtbH}
\eeq 
where $P_{L,R}=1/2(1\mp\gamma_5)$ are the chiral projector operators
and $V_{tb}$ is the corresponding CKM matrix element--henceforth
we set $V_{tb}=1$. 

The basic free parameters of our
analysis concerning the electroweak sector are contained in the
stop and sbottom mass matrices ($\tilde{q}=\tilde{t},\tilde{b}$):
\begin{equation}
{\cal M}_{\tilde{q}}^2 =\left(\begin{array}{cc}
 {\cal M}_{11}^2 & {\cal M}_{12}^2 
\\ {\cal M}_{12}^2 &{\cal M}_{22}^2 \,.
\end{array} \right)\,,
\label{eq:stopmatrix}
\end{equation} 
\begin{eqnarray}
{\cal M}_{11}^2 &=&M_{\tilde{q}_L}^2+m_q^2
+\cos{2\beta}(T^3_q-Q_q\,\sin^2\theta_W)\,M_Z^2\,,\nonumber\\
{\cal M}_{22}^2 &=&M_{\tilde{q}_R}^2+m_q^2
+Q_q\,\cos{2\beta}\,\sin^2\theta_W\,M_Z^2\,,\nonumber\\
{\cal M}_{12}^2 &=&m_q\, M_{LR}^q\nonumber\\
M_{LR}^{\{t,b\}}&=& A_{\{t,b\}}-\mu\{\cot\beta,\tan\beta\}\,. 
\end{eqnarray}
We denote by $m_{\tilde{q}_1}$ ($\tilde{q}=\tilde{t},\tilde{b}$) the
lightest stop/sbottom mass-eigenvalue.
For the sake of simplicity, we treat the sbottom mass matrix assuming 
that $\theta_b=\pi/4$, so that the two
diagonal entries are equal.
This is no loss of
generality, since $t\rightarrow W^+\,b$ turns out to be rather insensitive
to SUSY physics, and on the other hand
the feature that really matters for our calculation
on $t\rightarrow H^+\,b$
is that the off-diagonal element of the sbottom mass
matrix is non-vanishing, 
so that at high $\tan\beta$ it behaves like  
$m_b\,M_{\rm LR}^b\simeq -\mu\,\,m_b\tan\beta$. 
As for the stop mixing angle, it is
determined by the remaining input parameters.
Finally, in the SUSY strongly 
interacting sector, the
only new parameter is the gluino mass, $m_{\tilde{g}}$.

Another fundamental ingredient of our renormalization framework is the
definition of $\tan\beta=v_2/v_1$ beyond the tree-level.
The quantum analysis of the standard decay of the top 
quark, $t\rightarrow W^+\,b$, does not require renormalization of $\tan\beta$. 
Nonetheless, a full one-loop calculation of $t\rightarrow H^+\,b$
within the MSSM requires a prescription to renormalize that parameter.
At one-loop, $\tan\beta$ is not renormalized by SUSY-QCD, but of course
it is affected by the electroweak corrections.
Any definition will generate a
counterterm,
$
\tan\beta\rightarrow\tan\beta+\delta\tan\beta\,,
$
which depends on the specific renormalization condition.
There are many conceivable strategies. The ambiguity is 
related to the fact that this parameter is just a Lagrangian parameter 
and as such it is not a physical observable.
Its value beyond the tree-level is
renormalization scheme dependent.
For example, we may wish to define $\tan\beta$ in a 
process-independent (``universal'') way
as the ratio $v_2/v_1$ between the true VEV's after renormalization of the
Higgs potential. But, then, we have to take care of the induced
linear terms (``tadpoles''). 
In other words, we have to renormalize them away
by shifting the VEV's and the mass parameters,
\beqn 
v_i &\rightarrow & Z_{H_i}^{1/2} (v_i+\delta v_i)\,,\nonumber\\
m_i^2 &\rightarrow & Z_{H_i}^{1\over 2}\,(m_i^2+\delta m_i^2)\,,\nonumber\\
m_{12}^2 &\rightarrow & Z_{H_1}^{1\over 2}\,Z_{H_2}^{1\over 2}
\,(m_{12}^2+\delta m_{12}^2)\,,
\label{eq:shifts}
\eeqn
in the Higgs potential
\,\cite{Hunter}:
\beqn
V &=& m_1^2\,|H_1|^2+m_2^2\,|H_2|^2-m_{12}^2\,\left(
\epsilon_{ij}\,H_1^i\,H_2^j+{\rm h.c.}\nonumber\right)\\
&+&{1\over 8}(g^2+g'^2)\,\left(|H_1|^2-|H_2|^2\right)^2
+{1\over 2}\,g^2\,|H_1^{\dagger}\,H_2|^2\,,
\label{eq:potential}
\eeqn
where $Z_{H_i}=1+\delta Z_{H_i}\,(i=1,2)$ on 
eq.(\ref{eq:shifts}) are the Higgs doublet
field renormalization constants:
\beq
\left(\begin{array}{c}
H_1^0 \\ H_1^{-}
\end{array} \right)\rightarrow Z_{H_1}^{1/2}
\left(\begin{array}{c}
H_1^0 \\ H_1^{-}
\end{array} \right)\,,\ \ \ \ \ 
\left(\begin{array}{c}
H_2^{+} \\ H_2^0 
\end{array} \right)\rightarrow Z_{H_2}^{1/2}
\left(\begin{array}{c}
H_2^{+} \\ H_2^0 
\end{array} \right)\,.
\label{eq:ZH1H2}  
\eeq
However, that shifting procedure can be manifold\,\cite{ChankPok}. If only the VEV's
are shifted, and the $\overline{MS}$ scheme is adopted, then
$\tan\beta=\tan\beta (\nu)$
becomes a rapidly varying function of the renormalization scale $\nu$, which
renders it useless\,\cite{Gamberini}.
A consistent and more convenient choice to cancel the tadpole terms
in the renormalization of the Higgs potential is to
shift both the VEV's as well as the mass parameters in such a way that 
$\delta v_1/v_1=\delta v_2/v_2$. From eq.(\ref{eq:shifts}), this obviously
entails
\beq
{\delta\tan\beta\over\tan\beta}=
{1\over 2}\,\left(\delta Z_{H_2}-\delta Z_{H_1}\right)\,.
\label{eq:ddtan}
\eeq 
In practice,   
process-dependent terms are inevitable irrespective of the definition
of $\tan\beta$. In particular, the 
definition of $\tan\beta$ where
 $\delta v_1/v_1=\delta v_2/v_2$ will
also develop process-dependent contributions,
as can be seen by trying to relate the ``universal'' value
of $\tan\beta$ in that scheme with a physical quantity directly read off 
some physical observable. For instance, if $M_{A^0}$ 
is heavy enough, one may define $\tan\beta$ as follows:
\beqn
{\Gamma (A^0\rightarrow b\,\bar{b})\over
\Gamma (A^0\rightarrow t\,\bar{t})}&=&
\tan^4\beta\,{m_b^2\over m_t^2}\,\left(1-{4\,m_t^2\over
 M_{A^0}^2}\right)^{-1/2}\,\left[1+4\,\left({\delta v_2\over v_2}-
{\delta v_1\over v_1}\right)
\right.\nonumber\\
& &\left.+2\,\left({\delta m_b\over m_b} 
+\frac{1}{2}\delta Z_L^b +\frac{1}{2}\delta Z_R^b
-{\delta m_t\over m_t} 
-\frac{1}{2}\delta Z_L^t-\frac{1}{2}\delta Z_R^t\right)
 +\delta V\right]\,,
\label{eq:tanbeta3}
\eeqn
where we have neglected $m_b^2\ll M_{A^0}^2$, 
and $\delta V$ stands for the vertex
corrections to the decay processes $A^0\rightarrow b\,\bar{b}$ and 
$A^0\rightarrow t\,\bar{t}$.
Since the sum of the mass and wave-function renormalization
terms along with the vertex corrections is UV-finite, one can consistently choose
$\delta v_1/v_1=\delta v_2/v_2$ leading to eq.(\ref{eq:ddtan}).
Hence, deriving $\tan\beta$ from eq.(\ref{eq:tanbeta3}) 
unavoidably incorporates also some
process-dependent contributions. In general, process-dependent effects
can be rather important, as shown
in Ref.\cite{CGGJS}; and, therefore, they can substantially alter the numerical
comparison between the various $\tan\beta$ definitions presented in
Ref.\cite{YYamada}. 

In practice, one may eventually like to 
fix the on-shell renormalization condition on $\tan\beta$ in a more physical
way, viz. by relating it to some concrete physical observable, so that it is
the measured value of this observable that is taken as an input rather 
than the VEV's of the Higgs potential. 
Following this practical attitude, we choose as a physical observable 
the decay width of the charged Higgs boson into $\tau$-lepton and associated
neutrino:
\beq
\Gamma(H^{+}\rightarrow\tau^+\nu_{\tau})=
{\alpha m_{\tau}^2\,M_{H}\over 8 M_W^2 s_W^2}\,\tan^2\beta\,. 
\label{eq:tbetainput}
\eeq
This should be a very good
choice, since this decay is the leading decay mode of $H^+$ at high
$\tan\beta$, which is the regime where
 $t\rightarrow H^+\,b$ is competitive with $t\rightarrow W^+\,b$.
This definition produces the following counterterm\,\cite{CGGJS}:
\beq
{\delta\tan\beta\over \tan\beta}=
{\delta v\over v}-\frac{1}{2}\delta Z_{H^+}
+\cot\beta\, \delta Z_{HW}+ 
\Delta_{\tau}\,,
\label{eq:deltabeta}
\eeq    
where $v^2=v_1^2+v_2^2$, and
\beqn  
\delta Z_{H^{\pm}} &=& +\Sigma_{H^{\pm}}^{\prime}(M_{H^{\pm}}^2)\,,\nonumber\\
\delta Z_{HW}&=&{\Sigma_{HW}(M_{H^{\pm}}^2)\over M_W^2}\,,
\eeqn
are the 
field renormalization constants associated to $H^+$ and the $H^+-W^+$ mixing.
The term $\Delta_{\tau}$ stands for a cumbersome expression containing 
the full set of MSSM corrections to the decay
$H^+\rightarrow\tau^+\,\nu_{\tau}$. 

Any definition of $\tan\beta$ is
in principle as good
as any other; and, in spite of the fact that
the corrections themselves may drag along some dependence
on the choice of the particular definition,
the physical observables should not depend at all on that choice.
Notwithstanding, it can be a practical matter what definition to use
in a given context.   
For example, our definition of $\tan\beta$ given in eq.(\ref{eq:tbetainput})
should be most adequate to analyze the decay $t\rightarrow H^+\,b$ (and so for 
charged Higgs masses $M_{H^{\pm}}<m_t-m_b$) 
and large $\tan\beta$ since, then,
$H^+\rightarrow\tau^+\,\nu_{\tau}$ is the dominant decay of $H^+$,
whereas the definition based on
eq.(\ref{eq:tanbeta3}) requires also a large value of $\tan\beta$ (to avoid
an impractical suppression of the $b\,\bar{b}$ mode); moreover, in order to 
be operative, it also requires
a much heavier charged Higgs boson, for it turns out that 
$M_{H^{\pm}} \simeq M_{A^0}>2\,m_t$ when
the decay $A\rightarrow t\bar{t}$ is kinematically open in the MSSM.\,
If however $m_t+m_b<M_{H^\pm}<2\,m_t$, the definition
(\ref{eq:tanbeta3}) cannot be used whereas the alternative
definition (\ref{eq:tbetainput})
can again be useful, e.g. to analyze $H^+\rightarrow t\,\bar{b}$\,\cite{CGGJS2}.

Within our context, we use 
eq.(\ref{eq:deltabeta}) for $\delta\tan\beta/\tan\beta$ in order 
to compute the one-loop corrections to the decay
$t\rightarrow H^+\,b$.                                  
The relative importance of the process-dependent contributions
associated to our definition of $\tan\beta$ is studied
in Ref.\cite{CGGJS}. Here I shall not split any more the 
oblique and non-oblique contributions,
and shall present only the full result.
After explicit calculation of all the counterterms in our 
renormalization scheme, as well as of the full plethora of MSSM
corrections to the three-processes: $t\rightarrow W^+\,b$,
$t\rightarrow H^+\,b$ and $H^+\rightarrow\tau^+\,\nu_{\tau}$, we
are ready to present the outcome of our analysis. We do it in terms of the 
relative shift with respect to the corresponding tree-level
width $\Gamma^{(0)}$:
\beq
\delta={\Gamma_{W,H}-\Gamma^{(0)}_{W,H}\over \Gamma^{(0)}_{W,H}}
\equiv{\Gamma (t\rightarrow \{W^+,H^+\}\,b)-
\Gamma^{(0)}(t\rightarrow \{W^+, H^+\}\,b)
\over \Gamma^{(0)}(t\rightarrow \{W^+,H^+\}\,b)}\,,
\label{eq:pito}
\eeq
In what follows we understand that $\delta$ defined by eq.(\ref{eq:pito})
are the corrections relative to the tree-level width
$\Gamma^0\equiv\Gamma_{\alpha}^0$ in the $\alpha$-scheme.
The corresponding correction with
respect to the tree-level width in the $G_F$-scheme
is simply given by\,\cite{GJSH} 
\beq
\delta \left ({G_F}\right)=\delta^{MSSM}-\Delta r^{MSSM}\,,
\label{eq:alphagf1}
\eeq
where $\Delta r^{MSSM}$ was object of a particular
study \,\cite{Modern,MW} and therefore it can be easily incorporated, if necessary.
Note, however, that $\Delta r^{MSSM}$ is already tightly
bound by the present experimental data on $M_Z=91.1863\pm 0.0020\,GeV$ 
at LEP\,\cite{LEPEWWG} and the ratio
$M_{W}/M_Z$ in
$p\bar{p}$, which lead
to $M_W=80.356\pm 0.125\,GeV$. 
Therefore, even without doing the exact theoretical calculation
of $\Delta r$ within the MSSM, we
already know from
\beq
\Delta r^{exp}=1-{\pi\alpha\over \sqrt{2}\,G_F}
\,{1\over M_W^2\,(1-M_W^2/M_Z^2)}= 0.040\pm 0.018\,,
\label{eq:deltar}
\eeq
that $\Delta r^{MSSM}$ must lie in this experimental interval. 
Since, as we have argued before, quantum effects in the MSSM do preserve the DT,
we expect (and we know\,\cite{Modern,MW}) that for the present
values of the sparticle masses 
$\Delta r^{MSSM}\leq\Delta r^{SM}\leq\Delta r^{exp}$.

We plot in Figs.1a-1d the results of the SUSY corrections to
$t\rightarrow W^+\,b$ within the framework of the MSSM
\footnote{These results update the analyses
presented in Refs.\cite{GJSH,DJJHS}.}.
For a better understanding of the various types of corrections, 
we introduce the following definitions:
\begin{itemize}
\item{(i)} The supersymmetric electroweak (SUSY-EW)
contribution from genuine ($R$-odd) sparticles,  i.e. 
from sfermions (squarks and sleptons), charginos and neutralinos:
$\delta_{SUSY-EW}$. 
\item{(ii)}  The electroweak contribution from 
non-supersymmetric ($R$-even) 
particles: $\delta_{EW}$. It is composed of two distinct types
of effects, namely, those from the MSSM Higgs 
and Goldstone bosons (collectively called
``Higgs'' contribution, and denoted $\delta_{H}$) plus
the oblique  SM effects\,\cite{BSH}
from conventional fermions ($\delta_{SM}$):
\beq
\delta_{EW}=\delta_{H}+\delta_{SM}\,.
\eeq
The remaining non-supersymmetric electroweak effects
are subleading and are neglected.
\item{(iii)} The strong supersymmetric QCD (SUSY-QCD) contribution
from squarks and gluinos: $\delta_{SUSY-QCD}$. 
\item{(iv)} The total supersymmetric contribution from sparticles:
\beq
\delta_{SUSY}=\delta_{SUSY-QCD}+\delta_{SUSY-EW}\,.
\eeq
\item{(v)}
The standard QCD correction from quarks
and gluons: $\delta_{QCD}$.
\item{(vi)}
The total MSSM contribution, $\delta_{MSSM}$,
namely, the net sum of all the previous contributions:
\beq
\delta_{MSSM}=\delta_{SUSY}+
\delta_{EW}+\delta_{QCD}.
\label{eq:individ}
\eeq
\end{itemize}  
We see from Fig.1a that even for very large (or very small) 
values of $\tan\beta$ -- within the perturbative range (\ref{eq:tanbeta1}) --  
the SUSY quantum corrections to $t\rightarrow W^+\,b$ are rather
tiny, already in the
$\alpha$-scheme. In the $G_F$-scheme, the corrections are still smaller,
for $\delta$ in the $\alpha$-scheme is comparable to $\Delta r$
and there is a cancellation between the two terms in eq.(\ref{eq:alphagf1}).
We point out that the SUSY-QCD effects on $t\rightarrow W^+\,b$
are basically insensitive to $\tan\beta$ (the only
dependence being in the masses of the squarks). 
Thus the evolution of $\delta_{SUSY}$ with $\tan\beta$ in Fig.1a is 
essentially due to the electroweak component.
Unfortunately, $\delta_{SUSY}$
stays all the time within a few percent for $\tan\beta$ values in the range
(\ref{eq:tanbeta1}) and typical sparticle masses of ${\cal O}(100)\,GeV$.
The precise dependence on the squark and gluino masses is detailed in
Figs.1b-1d. For the sake of better clarity, in Fig.1d we have split up
$\delta_{SUSY}$ into the $\delta_{SUSY-QCD}$
and $\delta_{SUSY-EW}$ components. In this figure,
we have also included the standard QCD correction 
(dotted horizontal line);  $\delta_{QCD}$
reaches $-10\%$ and stays basically constant in the present 
experimental $m_t$ range (i.e. $m_t=175\pm 6\,GeV$).
Moreover, as it is plain from Fig.1d, $\delta_{QCD}$
is generally  dominant over the 
SUSY-QCD and the SUSY-EW correction. The transient dominance of the
SUSY-QCD  corrections near $m_{\tilde{g}}=75\,GeV$ 
(see the downward spike in Fig.1d) is related to
a threshold effect on the two and three point functions
at $m_{\tilde{g}}=m_t-m_{\tilde{t}_1}$. Similar threshold 
behaviours are recorded in Figs.1b and 1c involving other sparticle masses.

The smallness of the SUSY corrections to  $t\rightarrow W^+\,b$
may be somewhat surprising, and it was not obvious
a priori, due to the appearance of enhanced top and bottom
quark Yukawa couplings
(\ref{eq:Yukawas}) scattered over the plethora of Higgs-quark-quark 
and chargino-quark-squark diagrams involved in the calculation.
Already in the pure SM context\,\cite{TopSM},
where one could also expect
large electroweak corrections of order $\alpha_W m_t^2/M_W^2$, the
actual corrections turn out to be rather small. Specifically,
they are positive and stay within $+(4-5)\%$ 
in the $\alpha$-scheme, and in a narrow interval
around $+1.7\%$ in the
$G_F$-scheme. These results hold for the present values of the top quark
mass, the variation with the SM Higgs mass being very mild. 
We point out that,
due to the SUSY constraints, 
the extra Higgs effects from the two-doublet Higgs sector
of the MSSM  (relatively to the SM)
are very small, typically one order of magnitude
smaller than the SM Higgs effects \,\cite{GRHoang}.

Therefore, since in the end the full Higgs effects
are negligible
we may set, in good approximation,
\beq 
\delta_{EW}\simeq\delta_{SM}\simeq\hat{\Sigma}_W^{\prime SM}(M_W^2)
\simeq -{\hat{\Sigma}^{SM}_W(0)\over M_W^2}=\Delta r^{SM}\,,
\label{eq:oblique}
\eeq
where we have used the fact that $\Delta r$ constitutes the bulk of 
the renormalized wave-function factor
$\hat{\Sigma}_W'(M_W^2)$ associated to the 
$W$-field. Indeed, the unrenormalized factor ${\Sigma}_W'(M_W^2)$ does 
not contain the leading SM
fermionic contributions. These are
brought along by the corresponding counterterm
$\delta Z_2^W$:
\beq
\hat{\Sigma}_W'(M_W^2)={\Sigma}_W'(M_W^2)-\delta Z_2^W\,,
\label{eq:rensigma}
\eeq
with\footnote{The notation is as in eqs. (27), (31) and (32) of Ref.\cite{GJSH}.}
\beqn
\delta Z_2^W & = &
\left.  {\Sigma_{\gamma}(k^2)\over k^2}\right|_{k^2=0}+2{c_W
\over s_W}\,{\Sigma_{\gamma Z}(0)\over M_Z^2}
+{c_W^2\over s_W^2}
\left({\delta M_Z^2\over M_Z^2}-{\delta M_W^2\over M_W^2}\right)\nonumber\\
&\simeq & -\Delta \alpha+{c_W^2\over s_W^2}\,\delta\rho+...
\simeq -\Delta r+...\,,
\label{eq:approx}
\eeqn
where we have formally neglected UV-divergent terms which cancel in the full
expression (\ref{eq:rensigma}).  
Similarly, we have used the formal approximation
\beq
\delta M_{W,Z}^2=
-\Sigma_{W,Z}(M_{W,Z}^2)\simeq -\Sigma_{W,Z}(0)+... 
\eeq
(where we dismiss terms other than leading fermionic contributions)
and the standard formulae
\beqn
\Delta r&=&\Delta\alpha-{c_W^2\over s_W^2}
\,\delta\rho+(\Delta r)_{rem.}\,,\nonumber\\
\Delta\alpha &=& \left.{\Sigma_{\gamma}(k^2)\over k^2}\right|_{k^2=M_Z^2}
-\left.{\Sigma_{\gamma}(k^2)\over k^2}\right|_{k^2=0}\,,\nonumber\\
\delta\rho &=&{\Sigma_W (0)\over M_W^2}-{\Sigma_Z (0)\over M_Z^2}\,,
\eeqn
which we consider within the same leading fermionic approximation.
The oblique contribution (\ref{eq:oblique}) cancels out
in the $G_F$-scheme -- Cf. eq.(\ref{eq:alphagf1}) -- but, 
as mentioned before, in the SM one is still left with
a non-oblique remainder of about $ +1.7\%$ which is basically unaltered
by the extra Higgs effects of the MSSM. 

From the foregoing we have
$\delta_{SM}\simeq\Delta r^{SM}\simeq +(3-4)\%$. However, by inspection
of  Fig.1a we see that typically the SUSY correction reads
$\delta_{SUSY}=-(2-4)\%$, i.e. it is negative and may cancel to a large extent
against the previous SM contribution. Thus 
the upshot is that 
\beq
\delta_{MSSM}\simeq \delta_{SUSY}+\delta_{SM}+\delta_{QCD}\simeq \delta_{QCD}\,.
\eeq
In other words, the total MSSM correction to $\Gamma (t\rightarrow W^+\,b)$
is mostly the standard QCD correction.
For this reason we have not plotted
the total $\delta_{MSSM}$ in Fig.1 but only the relevant SUSY part.
At the end of the day, the net SUSY quantum signature potentially hidden
behind the conventional top quark decay is fairly disappointing. 
This scanty result is reminiscent of
the analogue one with $Z$-boson physics mentioned above, the
``problem'' being that in both cases the original vertex is a
gauge boson-fermion-fermion interaction where the SUSY quantum
effects are doomed to be very small\footnote{For instance, it is well-known 
that the ``$R_b$ anomaly'', in its worse episode, could not be
completely accounted for in the MSSM, not
even pushing the bottom quark Yukawa coupling up to the maximum allowed value
by perturbation theory\,\cite{GJS2}. And to cure that anomaly it only required an
additional SUSY correction to the $Z\rightarrow b\,\bar{b}$ partial width 
of less than $+3\%$, which the MSSM could hardly afford!. 
As expected, the size (in absolute value) of the maximum SUSY correction demanded
to $\Gamma(Z\rightarrow\,b\,\bar{b})$  
is of the same order, though obviously a bit smaller, than the 
maximum SUSY effect found on $\Gamma(t\rightarrow W^+\,b$).}.

The case with the decay  $t\rightarrow H^+\,b$ is quite another story,
the corrections being much larger (Cf. Figs.2 and 3a-3d)
than in the decay  $t\rightarrow W^+\,b$.
Most important, the genuine SUSY part of these corrections can also be very large
and consequently the total MSSM correction is highly modulated by the SUSY
components.
The only feature that $t\rightarrow H^+\,b$ has in common
with $t\rightarrow W^+\,b$ is that 
the total Higgs contribution ($\delta_H$)
is also very small (Fig.3a).

In Fig.2 we show the partial width $\Gamma (t\rightarrow H^+\,b)$
including all the MSSM effects. This quantity is a physical
observable and so is independent of our particular
renormalization scheme. 
In the same figure we have simultaneously plotted the tree-level width
and the conventional QCD-corrected width (without any sort of SUSY effect),
just to show that in the presence of virtual SUSY effects the physical 
value of that observable could be significantly different.
As it is plain from Fig.2 (see also Fig.3a), already without SUSY
the standard QCD effects are rather large and
have opposite sign to the SUSY effects.
Finally, in Fig.2 we have also 
included the MSSM-corrected partial width of the conventional
top quark decay, $\Gamma (t\rightarrow W^+\,b)$. As noted above,
for a typical choice of sparticle masses,
the SUSY part of the correction to that decay is very small,  
and cancels in part against the standard electroweak correction,
$\delta_{EW}\simeq\Delta r^{SM}$.
The evolution with $\tan\beta$ of $\Gamma (t\rightarrow W^+\,b)$ 
is very mild as compared
with that of $\Gamma (t\rightarrow H^+\,b)$.   

To appraise the relative importance of quantum SUSY physics 
on $\Gamma (t\rightarrow H^+\,b)$, 
in Fig.3a we chart the various types of individual MSSM corrections  
using the very same notation defined above for the
standard decay $t\rightarrow W^+\,b$. 
The dependence on the squark and gluino masses is
exhibited in Figs. 3b-3d.
A distinctive feature of these corrections
stands out immediately: viz.
they grow very fast with $\tan\beta$.
For example, from Fig.3a we read
$\delta_{\rm MSSM}\simeq +27\%$ for
$\tan\beta=35\simeq m_t/m_b$; and at $\tan\beta\simeq 50$, which
is the preferred value claimed by $SO(10)$ Yukawa coupling unification
models\,\cite{SO10},
the correction is already $\delta_{\rm MSSM}\simeq +55\%$.
These are truly large corrections which occur for
values of the sparticle masses of ${\cal O}(100)\,GeV$. 
The evolution
with $m_{\tilde{b}_1}$ and $m_{\tilde{t}_1}$ shows a slow
decoupling (Figs.3b,c) while the dependence on $m_{\tilde{g}}$ is such that,
locally, the SUSY-QCD corrections 
slightly increase  with $m_{\tilde{g}}$ (Cf. Fig.3d)
and eventually decouple (not shown). 
Quite remarkably, the decoupling rate turns out to be
so slow that one may reach $m_{\tilde{g}}\sim 1\,TeV$ 
without yet undergoing dramatic suppression.

It is patent from Figs.2-3 that, after
including the MSSM radiative corrections, the
branching ratio of the charged Higgs mode can be of order $50\%$ for
$\tan\beta$ near a typical $SO(10)$ point
$\tan\beta= 45-50$. 
To be sure, such large values of the branching ratio of the Higgs mode
are not in contradiction with the known data on top
decays at the Tevatron, as explained in Ref.\cite{CGGJS}. In contrast,
with only standard QCD effects, the branching ratio of $t\rightarrow H^+\,b$
could barely reach $20\%$ for the same values of the parameters. It is 
thus crystal clear that
if those SUSY quantum signatures are there, they could hardly be missed
in the next generation of experiments!. 
 
I shall not dwell on the complicated structure of the full one-loop
MSSM effects\,\cite{CGGJS}.
Nevertheless, a few remarks may rapidly convince
us that important corrections on the decay width of $t\rightarrow H^+\,b$
were, indeed, to be expected.
This can be understood from the counterterm structure of the
charged Higgs-fermion-fermion vertex, eq.(\ref{eq:LtbH}),
which is very different from
that of the gauge boson-fermion-fermion vertex. In fact, the
corresponding counterterm Lagrangian is the following:
\beq
\delta{\cal L}_{Hbt}={g\over\sqrt{2}\,M_W}\,H^-\,\bar{b}\left[
\delta C_R\ m_t\,\cot\beta\,\,P_R+
\delta C_L\ m_b\,\tan\beta\,P_L\right]\,t
+{\rm h.c.}\,,
\label{eq:LtbH2}
\eeq
with
\beqn
\delta C_R &=& {\delta m_t\over m_t}-{\delta v\over v}
+\frac{1}{2}\,\delta Z_{H^+}+\frac{1}{2}\,\delta Z_L^b+\frac{1}{2}
\,\delta Z_R^t
-{\delta\tan\beta\over\tan\beta}+\delta Z_{HW}\,\tan\beta\,,\nonumber\\
\delta C_L &=& {\delta m_b\over m_b}-{\delta v\over v}
+\frac{1}{2}\,\delta Z_{H^+}+\frac{1}{2}\,\delta Z_L^t+\frac{1}{2}
\,\delta Z_R^b
+{\delta\tan\beta\over\tan\beta}-\delta Z_{HW}\,\cot\beta\,,
\label{eq:countert}
\eeqn
where in our scheme $\delta\tan\beta$ is
given by eq.(\ref{eq:deltabeta}). 

Clearly, from eq.(\ref{eq:countert}) large SUSY effects
on $t\rightarrow H^+\,b$
are attainable at high $\tan\beta$, 
mainly because of the presence of the counterterm $\delta m_b/m_b$.
This counterterm receives large contributions from both SUSY-QCD (strong) and
SUSY-EW (electroweak) loops.
Formally, we have here the same one-loop 
threshold effect from
massive particles that one has to introduce in order to correct
the ordinary massless contributions 
(i.e. to correct the standard QCD running bottom quark mass, see below) in 
SUSY GUT models\,\cite{SO10}.
In these models,
a non-vanishing sbottom mixing may lead to important SUSY-QCD quantum effects 
on the bottom mass: $m_b= m_b^{GUT}+\Delta m_b$, where $\Delta m_b$ is 
proportional to 
$M_{LR}^b\rightarrow -\mu\tan\beta $ at sufficiently high
$\tan\beta$.
In our case, however, the bottom mass is an input parameter
for the on-shell scheme and
the effect obviously has a different physical meaning. 

Although the results presented in Figs. 2-3 include the full
correction from all possible one-loop diagrams
in the MSSM (see Ref.\cite{CGGJS} for a full list), it is illustrative
to pick up the leading contributions stemming from the $\delta m_b/m_b$
term\footnote{Previous partial
results can be found in the literature\,\cite{Previous}, but in no one of them
the leading effects were clearly recognized nor the
relative importance of the various contributions was fully assessed.}.
This is most easily done  
from the diagrams in the electroweak-eigenstate basis.
From mixed LR-sbottoms and gluino loops in Fig.4a we find
the finite term
\beqn
\left({\delta m_b\over m_b}\right)_{\rm SUSY-QCD} &=&
C_F\,{\alpha_s(m_t)\over 2\pi}\,m_{\tilde{g}}\,M_{LR}^b\,
I(m_{\tilde{b}_1},m_{\tilde{b}_2},m_{\tilde{g}})\nonumber\\
&\rightarrow & -{2\alpha_s(m_t)\over 3\pi}\,m_{\tilde{g}}\,\mu\tan\beta\,
I(m_{\tilde{b}_1},m_{\tilde{b}_2},m_{\tilde{g}}) \,,
\label{eq:dmbQCD}
\eeqn
where $C_F=(N_c^2-1)/2\,N_c=4/3$ is a colour factor.
The last result holds for sufficiently large $\tan\beta$.
We have introduced the positive-definite function 
\beq
I(m_1,m_2,m_3)=
{m_1^2\,m_2^2\ln{m_1^2\over m_2^2}
+m_2^2\,m_3^2\ln{m_2^2\over m_3^2}+m_1^2\,m_3^2\ln{m_3^2\over m_1^2}\over
 (m_1^2-m_2^2)\,(m_2^2-m_3^2)\,(m_1^2-m_3^2)}\,.
\label{eq:I123}
\eeq
As an aside, we observe that the so-called 
light gluino scenario would not favour
the charged Higgs decay of the top quark
since eq.(\ref{eq:dmbQCD}) vanishes for $m_{\tilde{g}}=0$. Still,
sizeable electroweak supersymmetric effects could be around. 

The bulk of the SUSY electroweak corrections is also buried
in $\delta m_b/m_b$. They are induced by
$\tan\beta$-enhanced Yukawa couplings of the type (\ref{eq:Yukawas}). 
Specifically, from loops involving mixed LR-stops 
and mixed charged higgsinos (Cf. Fig.4b), one finds:
\beqn
\left({\delta m_b\over m_b}\right)_{\rm SUSY-Yukawa} &=&
-{h_t\,h_b\over 16\pi^2}\,\, {\mu\over m_b}\,m_t\,M_{LR}^t
I(m_{\tilde{t}_1},m_{\tilde{t}_2},\mu)\nonumber\\
&\rightarrow &
-{h_t^2\over 16\pi^2}\,\mu\tan\beta\,A_t\,
I(m_{\tilde{t}_1},m_{\tilde{t}_2},\mu)\,,
\label{eq:dmbEW}
\eeqn
where again the last expression holds for large enough $\tan\beta$.

Similarly, the main source of process-dependent effects 
lies in the corrections generated by the $\tau$-mass counterterm,
$\delta m_{\tau}/ m_{\tau}$, and can be easily picked out in the 
electroweak-eigenstate basis (see Fig.4c-4d) much in the same way as
we did for the $b$-mass counterterm. There are, however, some differences,
as can be
appraised by comparing the diagrams in Figs.4a,b and c,d, where we see that
in the latter case the effect derives from diagrams
involving $\tau$-sleptons with gauginos or mixed gaugino-higgsinos.  
An explicit computation
of the leading pieces yields
\beqn
{\delta m_{\tau}\over m_{\tau}} &=&
{g'^2\over 16\pi^2}\,\mu\,M'\,\tan\beta\,
I(m_{\tilde{\tau}_1},m_{\tilde{\tau}_2},M')\nonumber\\
& + & {g^2\over 16\pi^2}\,\mu\,M\,\tan\beta\,
I(\mu,m_{\tilde{\nu}_{\tau}},M)\,,
\label{eq:dmtau12}
\eeqn
where $g'=g\,s_W/c_W$ and $M', M$ are the soft SUSY-breaking
Majorana masses
associated to the bino $\tilde B$ and winos $\tilde{W}^{\pm}$, respectively,
and the function $I(m_1,m_2,m_3)$ is again given by eq.(\ref{eq:I123}).

We may now ascertain the reason why $t\rightarrow W^+\,b$ and
LEP physics cannot
generate comparably large quantum SUSY signatures.  
The counterterm configuration associated to vertices involving
gauge bosons and conventional fermions\,\cite{BSH} does {\it not}
involve the term
$\delta m_b/m_b$ nor any similar structure, so that one cannot expect
enhancements of the type mentioned above. Therefore, SUSY-QCD corrections
to $t\rightarrow W^+\,b$ 
are not foreseen to be particularly significant in this case,
but just of order $\alpha_s(m_t)/4\,\pi$ without any enhancement factor.
This is borne out by the numerical analysis in Fig.1.
The only hope for gauge boson 
interactions with top and bottom quarks to develop sizeable radiative
corrections is to appeal to large non-oblique
corrections triggered by the Yukawa terms (\ref{eq:Yukawas}).
However, even in this circumstance the results are rather disappointing. For,
at large $\tan\beta\geq m_t/m_b$, the
bottom quark Yukawa coupling (the only relevant one in these conditions)
gives a contribution of order 
\beq
{\alpha_W\over 4\,\pi}\,{m_b^2\,\tan^2\beta\over M_W^2}\stackM 
{\alpha_W\over 4\,\pi}\,{m_t^2\over M_W^2}\,,
\label{eq:Yb}
\eeq
which is numerically very close to the strong contribution $\alpha_s (m_t)/4\,\pi$.
In contrast, the SUSY-QCD effect associated to (\ref{eq:dmbQCD}) is of order
\beq
C_F\,{\alpha_s\over 4\,\pi}\,\,\tan\beta\,,
\label{eq:dmb2}
\eeq
where we have approximated 
$I(m_{\tilde{b}_1},m_{\tilde{b}_2},m_{\tilde{g}})\simeq 1/(2\,\tilde{m}^2)$
in the limit where there is no large hierarchy between the sparticle masses,
i.e. all of them being of order $\tilde{m}$. 
The ratio between (\ref{eq:dmb2}) and (\ref{eq:Yb}) is
\beq
C_F\,\left({\alpha_s\over\alpha_W}\right)\,
\,\left({M_W\over m_t}\right)^2\,
\tan\beta= {\cal O}(1)\,\tan\beta\,.
\eeq
Consequently, in the entire high $\tan\beta$ regime below 
the perturbative limit (\ref{eq:tanbeta1}), the SUSY-QCD effects
on the $t\,b\,H^{\pm}$-vertex are expected to
be a factor of order $\tan\beta$
larger than the SUSY-QCD and 
Yukawa coupling effects on $t\,b\,W^{\pm}$ and  $b\,\bar{b}\,Z$.

Similar considerations apply for the SUSY electroweak corrections 
from (\ref{eq:dmbEW}). Even though they 
are subleading as compared to the strong supersymmetric
effects, they are generally larger than the electroweak 
supersymmetric effects
on gauge boson observables. As a matter of fact, they embody the
(non-negligible) $20\%$ residual MSSM contribution
left over after the SUSY-QCD and standard QCD corrections cancel out 
at $\tan\beta\stackM 30$ (see Fig.2a).

The dominant MSSM effects on $\Gamma_H$ are, by far, 
the QCD and SUSY-QCD ones. As noted above, even the ordinary
QCD effects\,\cite{CD} can be very large here.
At large $\tan\beta$ they are approximately given by
\beq
\delta_{QCD}=-\frac{2\, \alpha_s(m_t)}{\pi}\;\left(\frac{8\pi^2-15}{36}
            + \ln\frac{m^2_t}{m^2_b}\right)\simeq -62\%\,\
 \ \ (\tan\beta>>\sqrt{m_t/m_b}\simeq 6)\,.
\label{eq:QCD2}
\eeq 
The exact ${\cal O}(\alpha_s)$ formula gives slightly below $-60\%$, and 
for better accuracy one should also include renormalization group improvement.
The big log factor $\ln{m_t^2/m_b^2}$ is the necessary building block 
to construct the running b-quark mass
at the top quark scale, which is the energy scale of the process.
The QCD correction is of course negative since
$\alpha_s(q^2)$  is asymptotically free and so it
significantly decreases from $q^2=m_b^2$ to $q^2=m_t^2$.

We note that there can be a strong competition between QCD and
SUSY-QCD effects.
For $\mu<0$ (which in practice is the
only tenable possibility\,\cite{CGGJS}) $\delta_{QCD}$ and 
$\delta_{SUSY-QCD}$ have opposite signs 
(compare with the leading SUSY-QCD term, eq.(\ref{eq:dmbQCD}).
Therefore, there is a crossover point of the
two strongly interacting dynamics where the conventional QCD 
loops are fully cancelled by the SUSY-QCD loops.
This leads to the funny situation noted above, namely,
that the total MSSM correction is given by just the 
subleading, albeit non-negligible, electroweak supersymmetric 
contribution: $\delta_{\rm MSSM}\simeq \delta_{\rm SUSY-EW}$.
The crossover point occurs at $\tan\beta\stackM m_t/m_b$,
where $\delta_{\rm SUSY-EW}\stackM 20$.
For larger and larger $\tan\beta>m_t/m_b$,
the total (positive) MSSM correction grows very fast, since the
SUSY-QCD loops largely overcompensate the standard QCD corrections.
As a result, the net effect on the partial width
appears to be opposite in sign to what might
naively be ``expected'' (i.e. the QCD sign). This should leave an
indelible imprint on the quantum dynamics of the decay mode
$t\rightarrow H^+\,b$
which could be crucial to identify the SUSY nature of $H^\pm$.    
While a first significant test of these effects
could possibly be performed at the upgraded Tevatron, a
more precise verification would most likely
be carried out in future experiments 
at the LHC. 

A typical common set of inputs has been chosen
in Figs.2-3 such that the supersymmetric electroweak
corrections do reinforce the
strong supersymmetric effects (SUSY-QCD). For this set of inputs, the total MSSM
correction to the partial width of $t\rightarrow H^+\,b$ 
is positive for $\tan\beta> 20$ (approx.).
This is because the leading SUSY-QCD correction (at high $\tan\beta$)
is proportional to $\,-\mu$ (Cf. eq.(\ref{eq:dmbQCD})) and we are concentrating
our numerical analysis on just
the $\mu<0$ case.
 
Together with $A_t>0$, this yields $A_t\,\mu<0$,
which is a sufficient condition\,\cite{Ng} for the MSSM prediction of
$BR(b\rightarrow s\,\gamma)$ to be compatible with
experiment in the presence of a relatively light
charged Higgs boson  (as the one participating in the
top decay under study).
We recall that charged Higgs
bosons of ${\cal O}(100)\,GeV$ interfere constructively with the
SM amplitude and would render
a final value of $BR(b\rightarrow s\,\gamma)$ exceedingly high. 
Fortunately, this conclusion
can be circumvented in the MSSM since the alternative contribution from charginos
and stops tends to cancel the Higgs contribution provided that
$A_t\,\mu<0$. Furthermore, one must also
require relatively light values for the masses of the
lightest representatives of these sparticles, as well
as high values of $\tan\beta$\,\cite{Ng}; hence one is led to
a set of conditions 
which fit in with nicely to build up a favourable scenario
for the decay $t\rightarrow H^+\,b$. 

\vspace{1cm} 
\begin{Large}
{\bf 3. Direct $t\rightarrow {\rm SUSY}$ decays }
\end{Large}
\vspace{0.5cm}

Due to the large mass of the top quark,
there is plenty of phase space available
for two-body and multibody decays.
As for the direct decays of the top quark into two (R-odd) SUSY particles
the leading modes are the following\,\cite{Twobody,Guasch1}. If there is
a light gluino window, $m_{\tilde{g}}={\cal O}(1)\,GeV$, 
a pure SUSY-QCD decay is conceivable:
\beq
t\rightarrow \tilde{t}_1\,\tilde{g}\,,
\label{eq:tt1g}
\eeq
but, then, that top decay would be overwhelming
and should have been observed at the Tevatron. Therefore, 
since a light stop is recommended\,\cite{WdeBoer},
we shall consider that decay
unlikely.
On the SUSY electroweak side we have
some decays which cannot be ruled out yet: 
\beq
t\rightarrow \tilde{b}_a\,\chi^+_i\,,
\label{eq:tb1Psi1}
\eeq
and
\beq
t\rightarrow \tilde{t}_a\,\chi^0_{\alpha}\,.
\label{eq:tt1Psi0}
\eeq
Notice that several final states with chargino-neutralinos
could be kinematically
permitted among the possible ones ($a=1,2; i=1,2; \alpha=1,...,4$).
In Figs. 5a-5b we illustrate a case where some of these two-body
decays could be very important as compared to the SM decay. In particular,
the most favoured modes involve,
not the lightest, but the next-to-lightest chargino-neutralino, the reason
being that the latter are more higgsino-like and so have larger
Yukawa couplings of the form (\ref{eq:Yukawas}) than the former.

Be as it may, it could well happen 
that our most cherished SUSY two-body decays are not kinematically allowed
or are significantly suppressed in some regions of parameter space.
In this circumstance, some supersymmetric
three-body decays could still be significant.
Only very recently a systematic study of the direct top decays into three
(R-odd) SUSY particles has been made available in the literature, 
see Ref.\cite{Guasch1}. These $3$-body decays could be important not only
because they may lead 
to exotic, highly non-standard, signatures
(see Tables I and II of Ref.\cite{Guasch1}) but also
from the point of view of a future determination of the ${\it total}$ top
quark width, $\Gamma_t$. 

Indeed, for a consistent treatment of
the ``quantum signatures'' embodied in the observable $\Gamma_t$
at second order of perturbation theory (in the 
strong and electroweak gauge couplings) one should include the tree-level
contributions from all possible three-body decays
of the top quark in the MSSM. As it happens, the contribution of some
of these three-body decays turns out to be comparable to the largest SUSY quantum
effects on the standard top quark decay, $t\rightarrow W^+\,b$.

From the point of view of an {\sl inclusive} model-independent
measurement of  
the {\sl total} top-quark width, $\Gamma_t$, the
future $e^+\,e^-$ supercollider (NLC) 
should be a better suited machine. In an inclusive measurement,
all thinkable non-SM effects would appear on top
of the corresponding SM effects already computed in the
literature\,\cite{TopSM}. 
As shown in Ref.\cite{Fujii}, one expects to be able to measure the top-quark
width in $e^+\,e^-$ supercolliders at a level of $\sim 4\%$ on the basis of a
detailed analysis of both the top momentum distribution and the
resonance contributions to the forward-backward
asymmetry in the $t\bar{t}$ threshold region.

Although there are a roster of supersymmetric three-body decays of the
top quark that could be kinematically open in the light of 
the present sparticle bounds,
only a few can be of interest. In Ref.\cite{Guasch1}, we arrived at the
conclusion that the (potentially) most relevant modes are the following:
\beqn
&{\bf IV}.& t \rightarrow b\, \chi_{\alpha}^0\, \chi_i^+\nonumber\\
&{\bf VIII}.& t \rightarrow b\, \tilde{g}\, \chi_i^+\nonumber\\
&{\bf IX}.& t\rightarrow b\,\tilde{t}_b \, \bar{\tilde{b}}_a\nonumber\\
&{\bf X}.& t\rightarrow b\,\tilde{\tau}^+_a\, \tilde{\nu}_{\tau}\,,
\label{eq:processes}
\eeqn
which we assign the same numbering as in Ref.\cite{Guasch1}.
Let us denote by $\delta_i\ \ (i=IV,VIII,IX,X)$ the partial width of these
supersymmetric three-body decays of the top quark relative 
to the SM partial width, $\Gamma_{SM}\equiv \Gamma (t\rightarrow W^+\,b)$:
i.e. 
\beq
\delta_i={\Gamma (t\rightarrow 3\,\,{\rm body})\over
\Gamma_{SM}}\,.
\eeq
In Fig.5c,
$\delta_{IV}$ is studied as a function of $\tan\beta$ for parameter
values close to those used in Section 2. 
In this frame of inputs, Decay IV becomes more suppressed as
compared to the pay-off delivered in optimizing conditions\,\cite{Guasch1},
but still it can reach a few percent, and thus be non-negligible
as compared to the ordinary QCD corrections to the standard
top width $\Gamma(t\rightarrow W^+\,b)$. 
Decay VIII on the other hand, can 
yield $\delta_{VIII}=2\%$ (not shown) in similar
conditions, but it requires light gluinos.

As for the Decay IX, it is a very interesting three-body decay.
Nevertheless it can be relevant only if the $2$-body channel 
$H^+\rightarrow\tilde{t}\,\bar{\tilde{b}}$ is kinematically forbidden, 
otherwise the Higgs decay width becomes too large and it has a
dramatic suppression on the relevant $3$-body mode.
In Fig.5d we study the evolution of $\delta_{IX}$ as a function of
$\tan\beta$. Among the four amplitudes contributing to this decay, the
most relevant one is the Higgs mediated amplitude, since it contains the
Higgs-stop-sbottom coupling which can be very large. For parameter values
comparable to those in Section 2,  $\delta_{IX}$ can reach a few percent
and thus also compete with the quantum effects on $t\rightarrow W^+\,b$.
Decay X is similar to IX but it is numerically less significant.

Finally, we mention that one can envision situations where all
these $3$-body decays can be further enhanced\,\cite{Guasch1}. 
In particular, Decay IX is extremely sensitive to large values of $\mu$,
$A_b$ and $\tan\beta$, which in appropriate conditions could make it
competitive with any of the aforementioned two-body modes.

\vspace{1cm} 
\begin{Large}
{\bf 4. Hadronic Higgs decays in the MSSM }
\end{Large}
\vspace{0.5cm}

We can easily convince ourselves of the relevance 
of addressing the issue of the
width of a Higgs boson. Just notice that
if a heavy neutral Higgs is discovered and is found to have
a narrow width, it would certainly not be the SM Higgs, whilst
it could be a SUSY Higgs.
For, a heavy enough SM Higgs boson is expected to rapidly
develop a broad width through decays
into gauge boson pairs whereas the SUSY Higgs bosons cannot
in general be that broad since their couplings to gauge bosons
are well-known to be suppressed\,\cite{Hunter}.
In compensation, their couplings to
fermions (especially to heavy quarks)
can be considerably augmented.
Thus the width of a
SUSY Higgs should to a great extent be given by its hadronic width;
even so a heavy $H^0$ and $A^0$ is in general
narrower than a SM Higgs of the same mass.
Alternatively, if the discovered
neutral Higgs is sufficiently light that it cannot decay into gauge boson
pairs, its decay width into relatively heavy fermion pairs such as
$\tau^+\,\tau^-$, and especially into $b\,\bar{b}$, could be much larger than
that of the SM Higgs, because of
$\tan\beta$-enhancement of the fermion couplings\,\cite{Hunter}.
Hence, it becomes clear that the hadronic width may play a very
important role in the study of the MSSM higgses, already at the tree-level.

Let us now consider the Higgs decays most sensitive to our SUSY quantum
signatures.
If the charged Higgs mass turns out to be larger than $m_t-m_b$, 
and the top quark
decay $t\rightarrow H^+\,b$ is thus kinematically forbidden, still interesting
information could be extracted from virtual Higgs contribution to
$t\rightarrow b\,\tau\,\bar{\nu}_{\tau}$\,\cite{Guasch1}
plus SUSY corrections. However, for very large
Higgs masses, we could still move to LHC physics and try to extract
information from the real charged Higgs 
decay $H^+\rightarrow t\,\bar{b}$. Similarly, there are
two regimes of Higgs masses ($M_{A^0}<2\,m_t$ and $M_{A^0}>2\,m_t$)
where the neutral Higgs can decay into hadrons, basically 
 $\Phi^i\rightarrow q\,\bar{q}$
with $\Phi^i= A^0, h^0, H^0$ and $q=t,b$. In the MSSM, $h^0$ cannot
decay into $t\,\bar{t}$, but $H^0$ and $A^0$ can decay both into $b\,\bar{b}$
and perhaps also into $t\,\bar{t}$, depending on their masses. 
These $q\,\bar{q}$ modes are the dominant decays of the charged and
neutral Higgs particles in the MSSM\footnote{See Ref.\cite{Zerwas} for a
recent comprehensive study of the two-body Higgs decays at the
tree-level in the MSSM.}. 
Although a fully-fledged one-loop analysis of the MSSM effects on these
hadronic modes
is not available within one and the same renormalization framework,
the main features from
SUSY-QCD are known\,\cite{Ricard}-\cite{Dabelstein}\footnote{If the Higgs particles
of the MSSM can decay into real squarks, radiative corrections to the
partial widths can also be significant\,\cite{Bartl2}. However, the SUSY corrections
to the more conventional hadronic modes $H^+\rightarrow t\,\bar{b}$ and
$\Phi^i\rightarrow q\,\bar{q}$ remain large even if sparticles are heavy enough
that Higgs bosons cannot decay directly into them\,\cite{Ricard}-\cite{Bartl}.}.

We define for the Higgs decays a relative correction, $\delta$, similar to 
eq.(\ref{eq:pito}).
In Figs.6a-6b we deal with the strong
(gluino mediated) SUSY effects ($\delta_{\tilde{g}}$) on
$H^+\rightarrow t\,\bar{b}$\,\cite{Ricard}. We show
the dependence of $\delta_{\tilde{g}}$
on the higgsino-mixing parameter $\mu$ and on the lightest sbottom mass.
In Figs. 7a-7b we illustrate
the behaviour of the strong SUSY effects on the hadronic neutral modes, as
a function of $M_{A^0}$ and various $\tan\beta$. The other Higgs masses are
determined by the usual MSSM Higgs mass relations. Since we are not including
the electroweak supersymmetric effects, we have used the tree-level mass
relations.

In Fig.7 we have simultaneousy plotted the standard (gluon mediated) QCD 
corrections ($\delta_g$). The QCD effects for $H^+\rightarrow t\,\bar{b}$
are also important (see Fig.5 of Ref.\cite{Ricard}).
In all Higgs decays (charged and neutral)
large supersymmetric effects are obtained at high $\tan\beta$
which could significantly modulate the conventional QCD corrections. 
Also noticeable is the high sensitivity of these decays
to the higgsino-mixing parameter
$\mu$. We observe that for the $b\,\bar{b}$ final states, the
dominant part of these corrections again originates 
from the bottom mass counterterm, eq.(\ref{eq:dmbQCD}). 
For the $t\,\bar{t}$ final states, instead, the leading corrections
come from the vertex form factors; however,
in this case the partial width becomes smaller the higher is $\tan\beta$.
Finally, we emphasize that in general all these 
effects remain substantial ($>10-20\%$) even for all sparticle masses well
above the LEP 200 discovery range. 

While we have just shown the impact of the standard QCD
and SUSY-QCD corrections, 
it is legitimate to worry about the larger and far more complex
body of electroweak
quantum effects, especially those coming from enhanced Yukawa
couplings. For the neutral Higgs bosons, this issue has already been 
addressed in other renormalization frameworks\,\cite{ChankPok}, but
there is no corresponding study of $H^+\rightarrow t\,\bar{b}$. 
A full analysis within our renormalization scheme is under
 way and will be presented
elsewhere\,\cite{CGGJS2}. Our preliminary results indicate that the
electroweak supersymmetric corrections are not negligible but are
definitely subdominant, except in the (unlikely) light gluino scenario.

In spite of the fact that we are focusing on decay processes, 
we should not forget 
that the $t\,b\,H^{\pm}$-vertex, and similar
$\Phi^i\,q\,q$ neutral Higgs vertices, can also be involved
in the production mechanisms (see Fig.8) and undergo significant
renormalization by SUSY effects.
In hadron machines an actual measurement
of the hadronic partial widths and in general of the effective hadronic vertices
$t\,b\,H^{\pm}$ and
$\Phi^i\,q\,\bar{q}\, (q=t,b)$ should be feasible. 
Let us briefly remind of the five basic 
mechanisms for neutral
Higgs production in a hadron collider\,\cite{WorldS}.
They have been primarily described for the
the SM Higgs, $H^0_{SM}$, but can be straightforwardly extended
to the three neutral higgses, $\Phi^i$, of
any $2$HDM:
\begin{itemize}
\item{(i)} 
Gluon-gluon fusion: $g\,g\rightarrow \Phi^i$;
\item{(ii)} $WW(ZZ)$ fusion:
$q\,q\rightarrow q\,q\,\Phi^i$;
\item{(iii)}
 Associated $W(Z)$ production: $q\,\bar{q}\rightarrow W (Z)\,\Phi^i$;
\item{(iv)} $t\,\bar{t}$ fusion:
 $g\,g\rightarrow t\,\bar{t}\,\Phi^i$, and 
\item{(v)} $b\bar{b}$ fusion: 
 $g\,g\rightarrow b\,\bar{b}\,\Phi^i$.
\end{itemize}
It is well-known\,\cite{Georgi} that, in the SM,
$g\,g\rightarrow H^0_{SM}$ fusion provides the
dominant contribution over most of the accessible range.
Nevertheless, for very large (obese) SM Higgs mass ($M_{H^0_{SM}}>500\,GeV$),
the $WW(ZZ)$-fusion mechanisms eventually
takes over; the rest of the mechanisms are subleading, and in particular 
$b\,\bar{b}$ fusion is negligible in the SM.
Remarkably enough, this situation could drastically change 
in the MSSM.
For instance, whereas 
one-loop $g\,g$-fusion in the SM is dominated by a top quark in the 
loop, this is not always so in the MSSM where the new couplings 
turn out to enhance, at high $\tan\beta$,
the $b$-quark loops and make them fully competitive with
the top quark loops.
Another example:  $b\,\bar{b}$ fusion, which is negligible in the SM,
can be very important in the MSSM at large
$\tan\beta$. As a matter of fact, for
large enough $\tan\beta$,
the $b\,\bar{b}$-fusion cross-section can be larger than
that for any mechanism for producing a SM Higgs boson
of similar mass.

What about the quantum effects on these production vertices?.
The conventional QCD corrections to $g\,g\rightarrow H^0_{SM}$
are known to be large\,\cite{DSZ}.
A similar conclusion holds for an obese 
SM Higgs boson
produced at very high energies by means of the
$WW(ZZ)$-fusion mechanisms; here, again,
non-negligible radiative effects do appear
\,\cite{Marciano}. Therefore, the production cross-section
for $H^0_{SM}$ is expected to acquire valuable quantum corrections
both for light and  for heavy Higgs masses. This is not so
for the corresponding width.
In fact, only for a heavy SM Higgs, namely, with a mass above
the vector boson
thresholds, the corrections to its decay width can
be of interest; for a light SM Higgs, instead,
light enough that it cannot decay into gauge boson pairs, the decay width
is very small and thus the corresponding quantum effects
are of no practical interest.

In contradistinction to the SM case,
the hadronic vertices $H^{\pm}\,t\,b$ and
$\Phi^i\,q\,\bar{q}$
could be the most significant interactions for MSSM higgses
irrespective of the value of the Higgs masses. In fact, these vertices
can be greatly enhanced and 
large radiative corrections could further shape the effective structure
of these interactions, as for example in
the charged Higgs mechanisms sketched in Fig.8.
In some of these mechanisms a Higgs boson is produced in
association, but in some others (fusion processes)
the Higgs boson enters as a virtual particle.
Now, however different these production processes might be,
all of them are sensitive to the effective structure
of the $H^{\pm}\,t\,b$ and $\Phi^i\,q\,\bar{q}$ vertices. 
While it goes beyond the scope of this report
to compute the SUSY corrections to the production processes
themselves, we have at least faced the detailed analysis of a partial
decay width which involves one of the
relevant production vertices. 
In this way, a definite prediction is made on the
properties of a physical observable and, moreover,
this should suffice both
to exhibit the relevance of the SUSY quantum effects
and to  demonstrate the necessity to incorporate these corrections 
in a future, truly 
comprehensive, analysis of the cross-sections, namely, an analysis where
one would include the quantum effects on all the relevant
production mechanisms within the framework 
of the MSSM. Some steps in this direction have already been given 
\,\cite{Moretti}. 
For this reason I think that   
in the future a precise measurement of the various (single and double) top
quark production cross-sections\,\cite{Willenbrock} 
will be able to detect or to exclude
the $t\,b\,H^{\pm}$-vertex as well as
the vertices $\Phi^i\,q\,\bar{q}$
involving the neutral Higgs particles
of the MSSM and the third generation quarks $q=t,b$.

I conclude the discussion on Higgs physics with a remark on $e^+\,e^-$
colliders and super-colliders. 
Thus far we have mostly elaborated on the Higgs
strategies at hadron colliders, such as the Tevatron and especially the LHC.
Now of course, in a future NLC, searches for quantum SUSY signatures
should also be feasible provided that the following processes
can be handled\footnote{Other $e^+\,e^-$ Higgs boson
production processes providing complementary 
information involve e.g. radiation of Higgs bosons off of top quarks, 
$ZZ$ fusion or $\gamma\gamma$ collisions, 
which could also be sensitive to SUSY quantum contributions.}:
\beqn   
e^+\,e^-&\rightarrow & Z\,h^0(H^0)\,,\nonumber\\
e^+\,e^-&\rightarrow & A\,h^0(H^0)\,,\nonumber\\
e^+\,e^-&\rightarrow & H^+\,H^-\,.
\label{eq:NLC}
\eeqn
The observed cross-sections for these processes
are equal to the production cross-sections times the Higgs
branching ratios.  Hence,
in an $e^+\,e^-$ environment, one aims more at
a measurement of the various
branching ratios (or, more precisely: ratios of branching ratios)
of the fermionic Higgs decay modes rather than of the
partial widths themselves.
For instance, in a $e^+\,e^-$ machine 
one would naturally address the measurement of
\beq
BR(\Phi^i\rightarrow b\,\bar{b})/BR(\Phi^i\rightarrow \tau^+\,\tau^-)\,.
\label{eq:bbtautau}
\eeq
This observable should receive large SUSY-QCD corrections
if $\Phi^i\rightarrow b\,\bar{b}$ proves to be, as we have seen,
very sensitive to the strong supersymmetric effects.
Unfortunately, among the processes (\ref{eq:NLC})
only $e^+\,e^-\rightarrow  Z\,h^0$ should obviously be
accessible to a typical NLC of $\sqrt{s}=500\,GeV$, and maybe even to LEP $200$.
This is because $m_{h^0}<150\,GeV$ in the MSSM whereas the
other Higgs particles can be much heavier. 
If the latter is indeed the case, and so $M_{A^0}$ is considerably large ($> 400\,GeV$),
then also the SUSY corrections
to $h^0\rightarrow b\,\bar{b}$ -- which we could measure e.g. by means of the
the $e^+\,e^-$ observable (\ref{eq:bbtautau}) -- turn out to be
very small (see Fig.7a) and in that event 
nothing firm could be decided on whether a hypothetically observed $h^0$
with mass $m_{h^0}<150\,GeV$ 
would be a SUSY Higgs or just the SM Higgs.  
In contrast, as also seen in Fig.7a, the SUSY corrections to
$H^0\rightarrow b\,\bar{b}$ and
$A^0\rightarrow b\,\bar{b}$ remain much larger, especially at high $\tan\beta$,
so that the form factors associated to the vertices $(H^0,A^0)\,q\,q$ are 
sensitive to SUSY corrections 
in a much wider kinematical range, probably well within the reach of the LHC 
\footnote{Higgs bosons unaccessible to LHC, i.e.
much heavier than $1\,TeV$, could hardly be considered as 
natural objects in the MSSM.}. 
Thus,  unless all of the MSSM Higgs
particles happen to be relatively light, or the nominal 
NLC energy  turns out to be amply superior
to $M_{A^0}+M_{H^0}\simeq 2\,M_{A^0}$
-- a requirement entailing $\sqrt{s}>2\,TeV$ in order to cover the worst 
possible scenario among the admissible ones --
the ability of $e^+\,e^-$ super-colliders
to fully explore the MSSM Higgs sector would be seriously hampered. 
Still, if $\sqrt{s}\stackm1\,TeV$,
at least three out of the five possible Higgs final states
in eq.(\ref{eq:NLC}) could perhaps be accessible\footnote{For example,
the European NLC is nominally planned to cover
a range $\sqrt{s}=500-800\,GeV$\,\cite{Manel}, maybe extendable
up to $1\,TeV$.}. 

In general, hadron colliders should have in spite of their own limitations 
a wider covering and also prove
greater sensitivity to SUSY quantum signatures.
This is because, as repeatedly emphasized, the main hadronic
decay modes involve the same basic dynamics as in the production modes,
namely on shell and off-shell Higgs-quark-quark vertices, respectively,
so that all Higgs bosons with mass below $1\,TeV$ can be singly produced
in a hadron machine
and some or all of the production mechanisms are
greatly sensitive to SUSY virtual effects.

\vspace{0.75cm} 
\begin{Large}
{\bf 5. Conclusions }
\end{Large}
\vspace{0.5cm}

In this talk I have discussed the main top and Higgs decays in the MSSM
by emphasizing the role played by the quantum effects as a means
to discriminate the potential SUSY dynamics underlying those decays.
The impact of direct top decays into sparticles has also been assessed
and compared its relative importance with the SUSY quantum effects. 
While on the one hand the generally small SUSY effects found out on
the standard top quark decay, $t\rightarrow W^+\,b$, unfortunately suggest that
only a forlorn attempt could be made to patch up some SUSY physics out
of top quark decay dynamics, 
the potentially large SUSY effects unveiled on the alternative decay
$t\rightarrow H^+\,b$  
should on the other hand raise a message of
hope for SUSY physics in hadron colliders. 
This conclusion is borne out also by our discussion
on charged and neutral Higgs decay
processes and Higgs production mechanisms which could be highly sensitive to
supersymmetric quantum effects.

Incidentally, we point out that
our MSSM analysis of $t\rightarrow H^+\,b$, followed 
by $H^+\rightarrow\tau^+\,\nu_{\tau}$, could have consequences for
$\tau$-lepton physics at the Tevatron. 
In particular, the bounds on the parameters $(\tan\beta,M_{H^\pm})$
involved in the charged Higgs decay of the top quark
could be significantly affected by quantum SUSY signatures.
In fact, the identification of the decay $H^+\rightarrow\tau^+\,\nu_{\tau}$
could be a matter of measuring a departure from the universality prediction
for all lepton channels, in the sense that
the experimental signature for $t\bar{t}\rightarrow H^+\,H^-\,b\,\bar{b}$
would differ from that of $t\bar{t}\rightarrow W^+\,W^-\,b\,\bar{b}$ by
an excess of final states with two $\tau$-leptons and two b-quarks 
and large transverse  missing energy.

In practice, $\tau$-identification is possible at the
Tevatron and there are recent analyses in the
literature\,\cite{Conway,Roy} finding excluded regions in the
$(M_{H^{\pm}},\tan\beta)$-plane on the basis of unsuccessful 
Tevatron searches for $\tau$-lepton violations of universality.
However, these exclusion plots are not fully watertight and
must be revised\,\cite{Guasch2}
in the light of our MSSM results on $t\rightarrow H^+\,b$.
Moreover, they should also be complemented with the 
highly competing information
extracted from the recent calculation of
SUSY corrections to low-energy semileptonic
$B$-meson decays\,\cite{CJS} where, again, a
similar pattern of large SUSY quantum corrections is found
\footnote{By the same token, one should not 
conclude -- as in Ref.\cite{Roy} -- that the
high $\tan\beta$ solution \,\cite{GJS2} to the (remnant!) $R_b$ ``anomaly''
is strongly disfavoured, without re-computing ``{\`a} la SUSY''
the excluded portion of the
$(M_{H^{\pm}},\tan\beta)$-plane. In particular, since the high $\tan\beta$
approach to $R_b$ is extremely sensitive to the value of the CP-odd Higgs mass
around $60\,GeV$\,\cite{GJS2}, even a slight modification of the allowed region
in the $(M_{H^{\pm}},\tan\beta)$-plane induced by the SUSY quantum effects
could be crucial\,\cite{Roypriv}.}.  
 
Our general conclusion is extremely encouraging for Tevatron and LHC physics:
In view of the potentially large size and large variety of manifestations,
quantum effects on physical processes involving
$H^\pm\,t\,b$, $\Phi^i\,b\,\bar{b}$ and/or  $\Phi^i\,t\,\bar{t}$ interactions
could be the clue to the discovery of ``virtual'' Supersymmetry.

\vspace{0.5cm}

{\bf Acknowledgements}:

This is an expanded written version of various
talks given at SUSY 96, CRAD 96, the CERN Theory Division, 
the UB, the IFAE/UAB and also, though in fairly preliminary form, at
the Rencontres de Moriond/ Electroweak 
Interactions 1996. It contains
a personal review of work done with my collaborators on the
specific subject of SUSY quantum effects on top quark and Higgs boson physics. 
I wish to take advantage of this opportunity to express my gratitude to
all of them.
In particular, to Wolfgang Hollik 
for sharing his knowledge and expertise on quantum corrections;
and to my collaborators in Barcelona, namely 
Toni Coarasa, David Garcia, Jaume Guasch
and Ricardo Jim{\'e}nez (the bulk of the ``IFAE SUSY Task Force'')  for 
many enjoyable discussions and for their 
invaluable help in the preparation of this manuscript.  
I am also thankful to
R. Mohapatra and to S. Jadach,  for their invitations
to SUSY 96 and CRAD 96, respectively;  to J.M. Fr{\`e}re for his invitation
to the Rencontres de Moriond 1996; and to the CERN Theory division,  particularly
to M. Carena and C.E.M. Wagner, for inviting me at CERN. 
This work has been partially supported by CICYT 
under project No. AEN95-0882.


\vspace{1cm}
\begin{center}
\begin{Large}
{\bf Figure Captions}
\end{Large}
\end{center}
\begin{itemize}
\item{\bf Fig.1} Total SUSY correction $\delta_{SUSY}$
(strong and electroweak) to the standard top quark decay,
$t\rightarrow W^+\,b$, for given values of the other
parameters. In the figure, $A_t=A_b\equiv A$.
Evolution of $\delta_{SUSY}$ with: (a) $\tan\beta$;
(b) the lightest sbottom mass, $m_{\tilde{b}_1}$; (c) the lightest 
stop mass, $m_{\tilde{t}_1}$; (d) the gluino mass, $m_{\tilde{g}}$. In (d)
we have split up the SUSY correction into SUSY-EW and SUSY-QCD effects and
have also included the standard QCD correction (horizontal dotted line).
As a rule, and unless stated otherwise, 
the parameters not explicitly defined in a figure
have common values with the other figures. 
The masses of the top and bottom
quarks are  $m_t=175\,GeV$ and $m_b=5\,GeV$, respectively.

\vspace{0.5cm}

\item{\bf Fig.2} The partial widths for the two decays
 $t\rightarrow W^+\,b$ and $t\rightarrow H^+\,b$ 
including all MSSM effects, versus $\tan\beta$, for a common
set of input parameters. We have simultaneously plotted
the tree-level width and the QCD-corrected width (without
SUSY effects) of the
decay $t\rightarrow H^+\,b$. For the standard decay,
$t\rightarrow W^+\,b$, the SUSY correction is very small and exhibits
a very mild evolution with $\tan\beta$ for the typical set
of parameters given. 

\vspace{0.5cm}

\item{\bf Fig.3} As in Fig.1, but for the MSSM corrections to
the charged Higgs decay of the top quark, $t\rightarrow H^+\,b$.
Particularly, in (a) we have detailed the SUSY-EW, standard EW,
SUSY-QCD, standard QCD, and total MSSM contributions.

\vspace{0.5cm}

\item{\bf Fig.4} Leading SUSY-QCD (a) and SUSY-EW (b) 
 contributions to $\delta m_b/m_b$ in
the electroweak-eigenstate basis. Similarly, (c) and (d)
show two independent sources of leading SUSY-EW 
contributions to $\delta m_{\tau}/m_{\tau}$.

\vspace{0.5cm}
  
\item{\bf Fig.5} (a) Partial width of the $2$-body SUSY decay 
$t\rightarrow \tilde{b}_a\,\chi^+_i$, as a function of $\tan\beta$,
relative to that of the SM decay $t\rightarrow W^+\,b$;
(b) As in (a), but for the decay 
$t\rightarrow \tilde{t}_a\,\chi^0_{\alpha}$;
(c) Partial width of the $3$-body SUSY decay
$t\rightarrow b\, \chi_{\alpha}^0\, \chi_i^+$, 
as a function of $\tan\beta$,
relative to that of the SM decay $t\rightarrow W^+\,b$;
(d) As in (c), but for the decay  
$t\rightarrow b\,\tilde{t}_b \, \bar{\tilde{b}}_a$. 
In all cases, the results are given for the
kinematically allowed mass eigenvalues 
($a,b=1,2; i=1,2; \alpha=1,...,4$)
compatible with the given set of inputs.

\vspace{0.5cm}

\item{\bf Fig.6}  Gluino-mediated SUSY-QCD corrections ($\delta_{\tilde{g}}$) 
to the partial width of the top quark decay of a charged Higgs boson,
$H^+\rightarrow t\,\bar{b}$, for fixed values of the other
parameters. Evolution of $\delta_{\tilde{g}}$ with:
(a) the higgsino-mixing parameter, $\mu$; 
(b) the lightest sbottom mass, $m_{\tilde{b}_1}$.

\vspace{0.5cm}

\item{\bf Fig.7}  Standard QCD ($\delta_g$) and
SUSY-QCD ($\delta_{\tilde{g}}$) corrections 
to the hadronic partial widths
 $\Gamma (\Phi^i\rightarrow q\,\bar{q})\ (q=b,t)$
of the neutral Higgs bosons:
(a) Dependence of $\Gamma(\Phi^i\rightarrow b\,\bar{b})$ 
on the CP-odd Higgs mass, $M_{A^0}$, for fixed
values of $\tan\beta$. The other Higgs masses are determined by the
usual mass relations in the MSSM Higgs sector.
(b) As in (a), but for 
the $\Gamma(A^0\,H^0\rightarrow t\,\bar{t})$. Here $A_t=400\,GeV$.

\vspace{0.5cm} 

\item{\bf Fig.8}  Typical diagrams for top quark and 
charged Higgs production
in hadron colliders involving the relevant $t\,b\,H^{\pm}$-vertex.
Similar diagrams can be deviced for the neutral Higgs boson 
$\Phi^i\,q\,q$-vertices ($q=t,b$) -- see Ref.\cite{ToniRicard}.

\end{itemize}

\newpage
\pagestyle{empty}

\vspace{-1cm}
\begin{tabular}{cc}
\epsfig{file=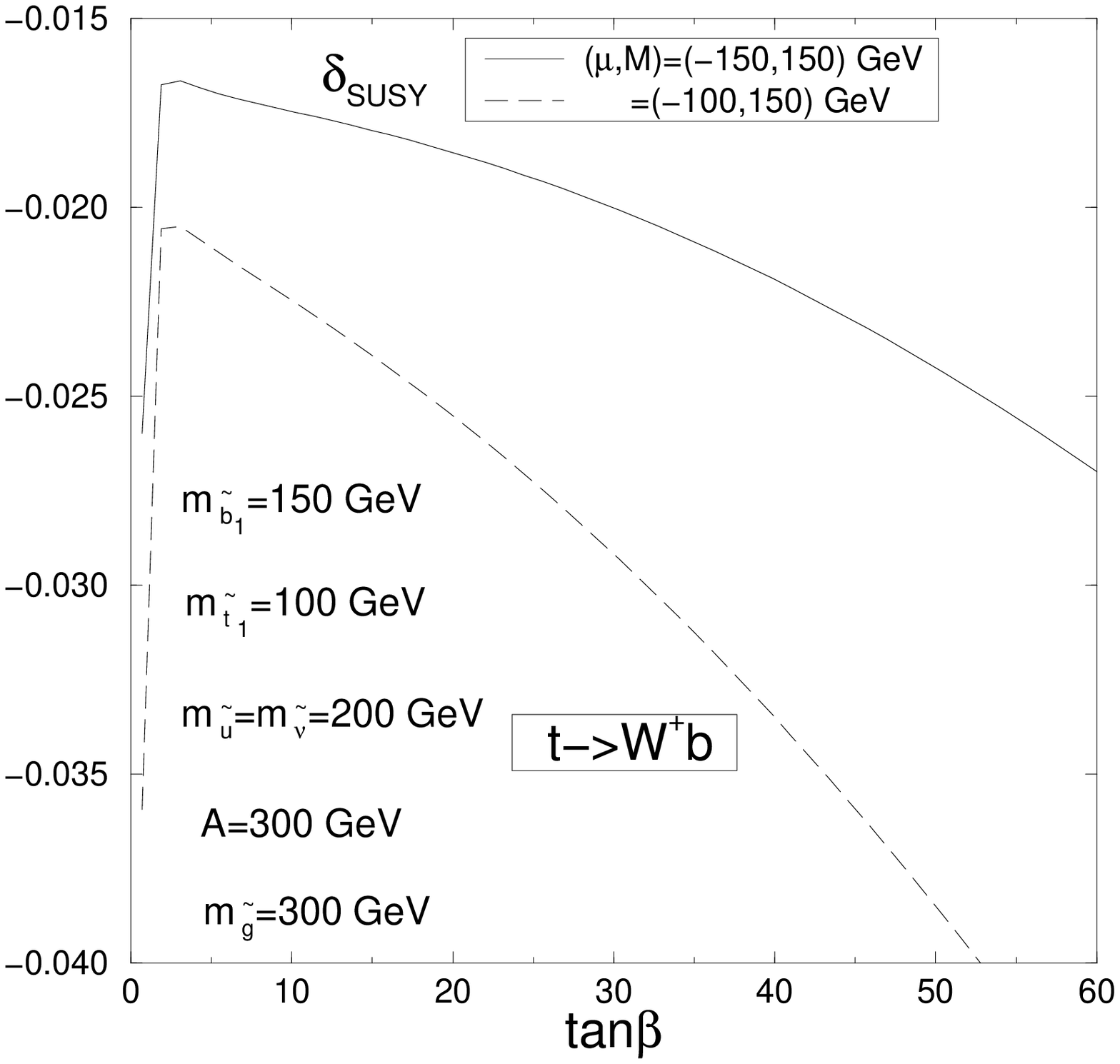,height=8cm,angle=-90}   &
\epsfig{file=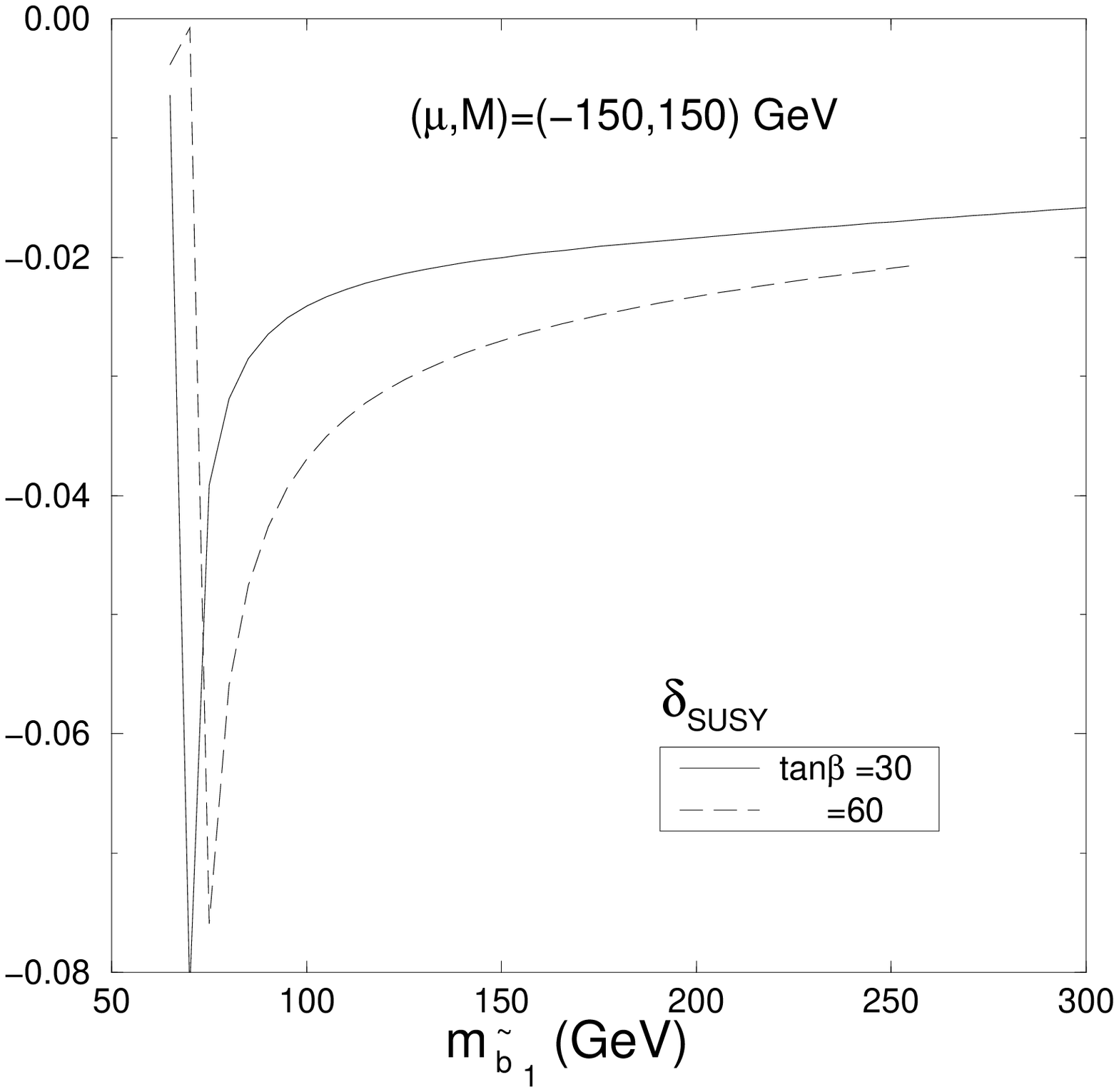,height=8cm,angle=-90} \\
(a) & (b) \\
\epsfig{file=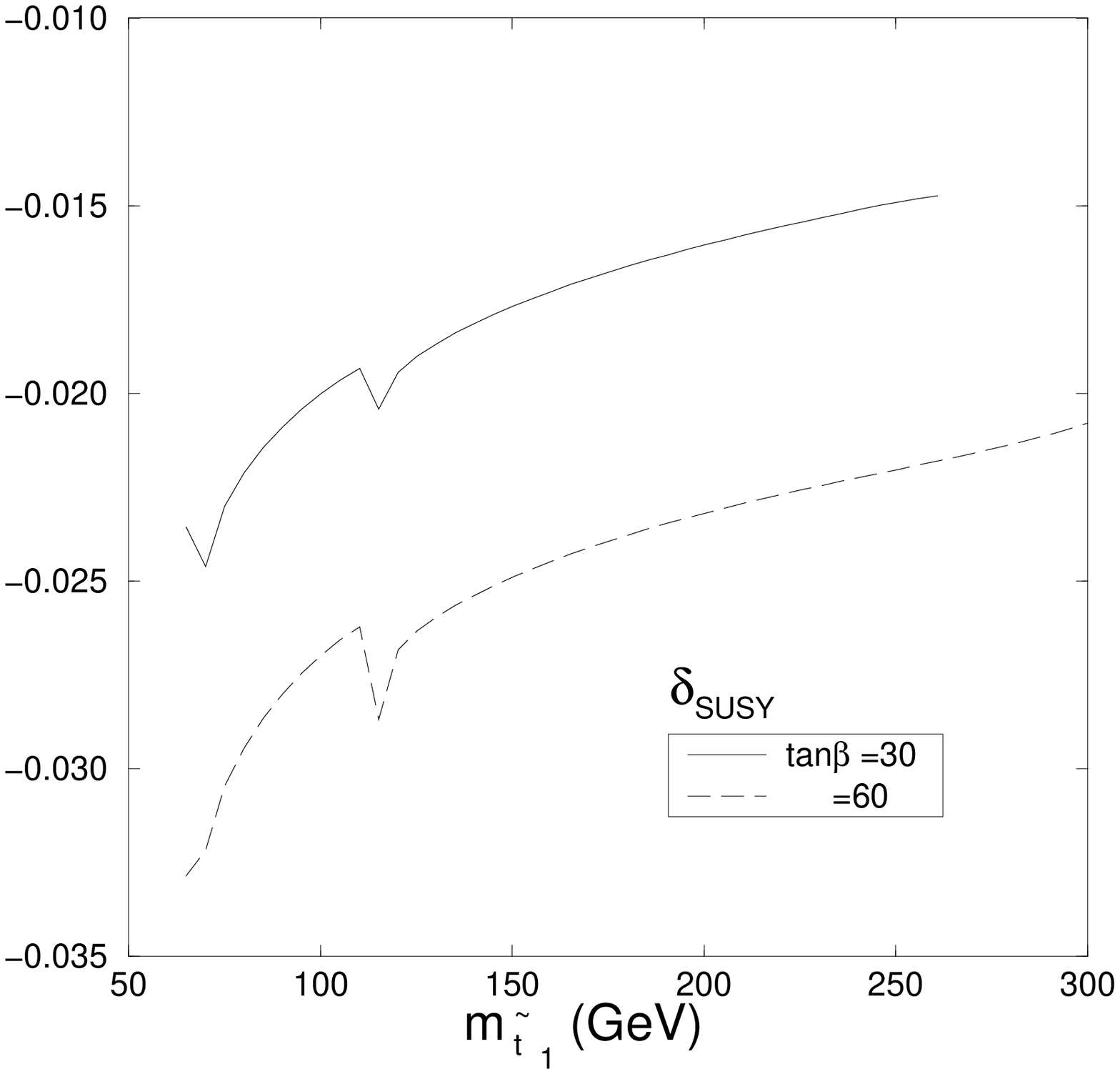,height=8cm,angle=-90} &
\epsfig{file=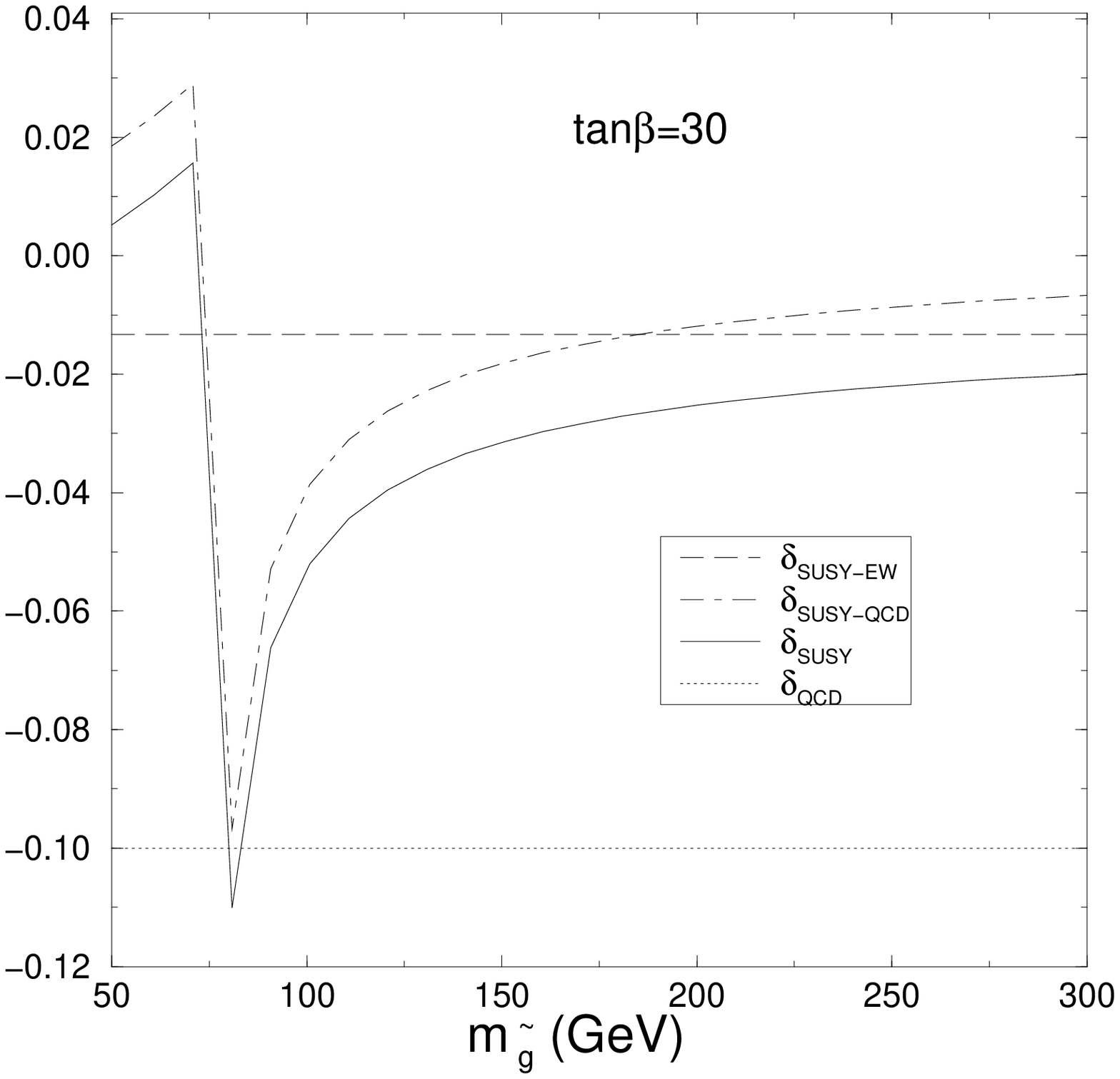,height=8.3cm,angle=-90} \\
(c) & (d)
\end{tabular}
\begin{center}
  {\Huge Fig.1}
\end{center}

\centerline{
\mbox{\epsfig{file=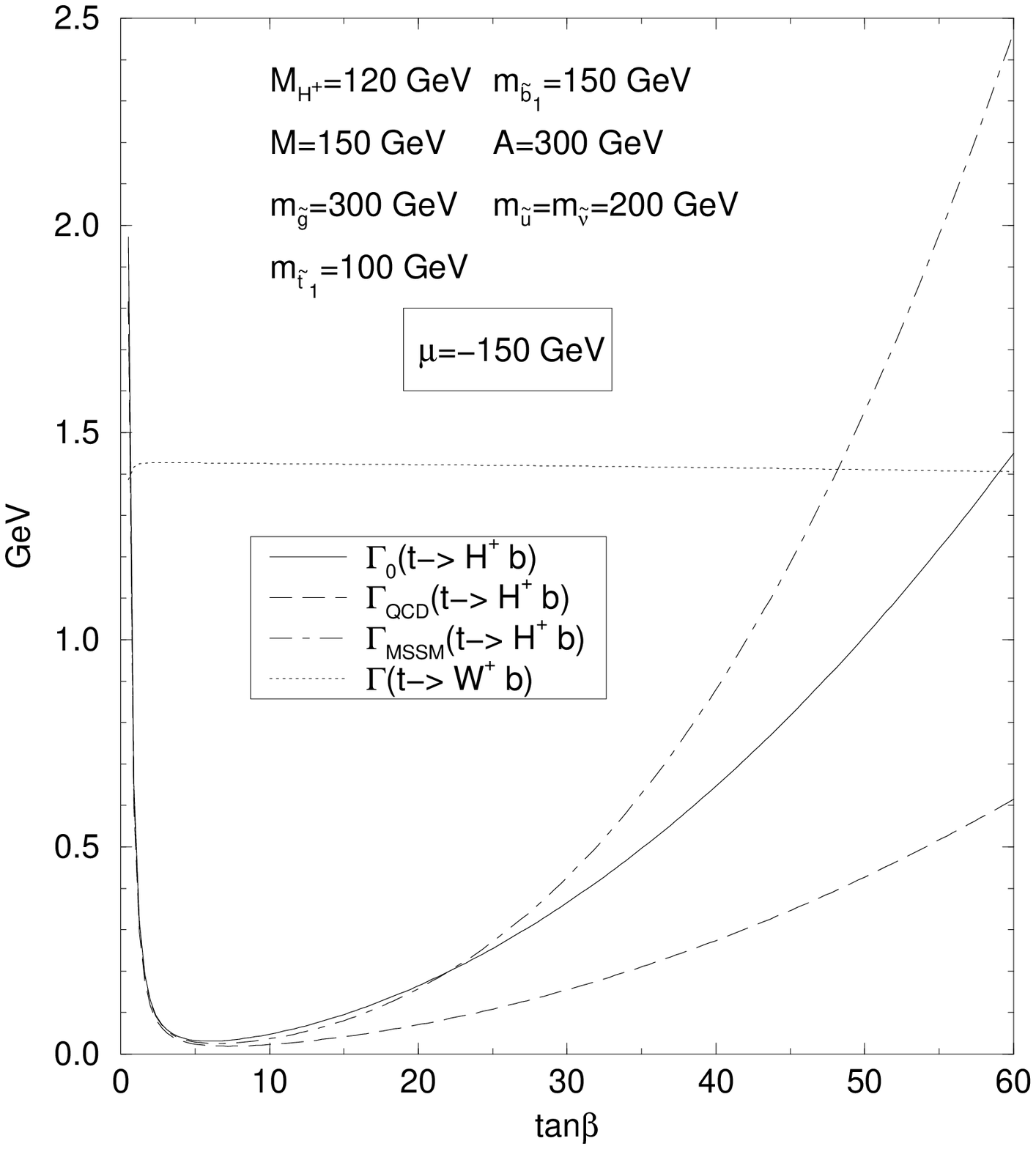,width=18cm}}
}
\begin{center}
{\Huge Fig.2}
\end{center}
\newpage

\centerline{
\begin{tabular}{c@{\hspace{1cm}}c}
\mbox{\epsfig{file=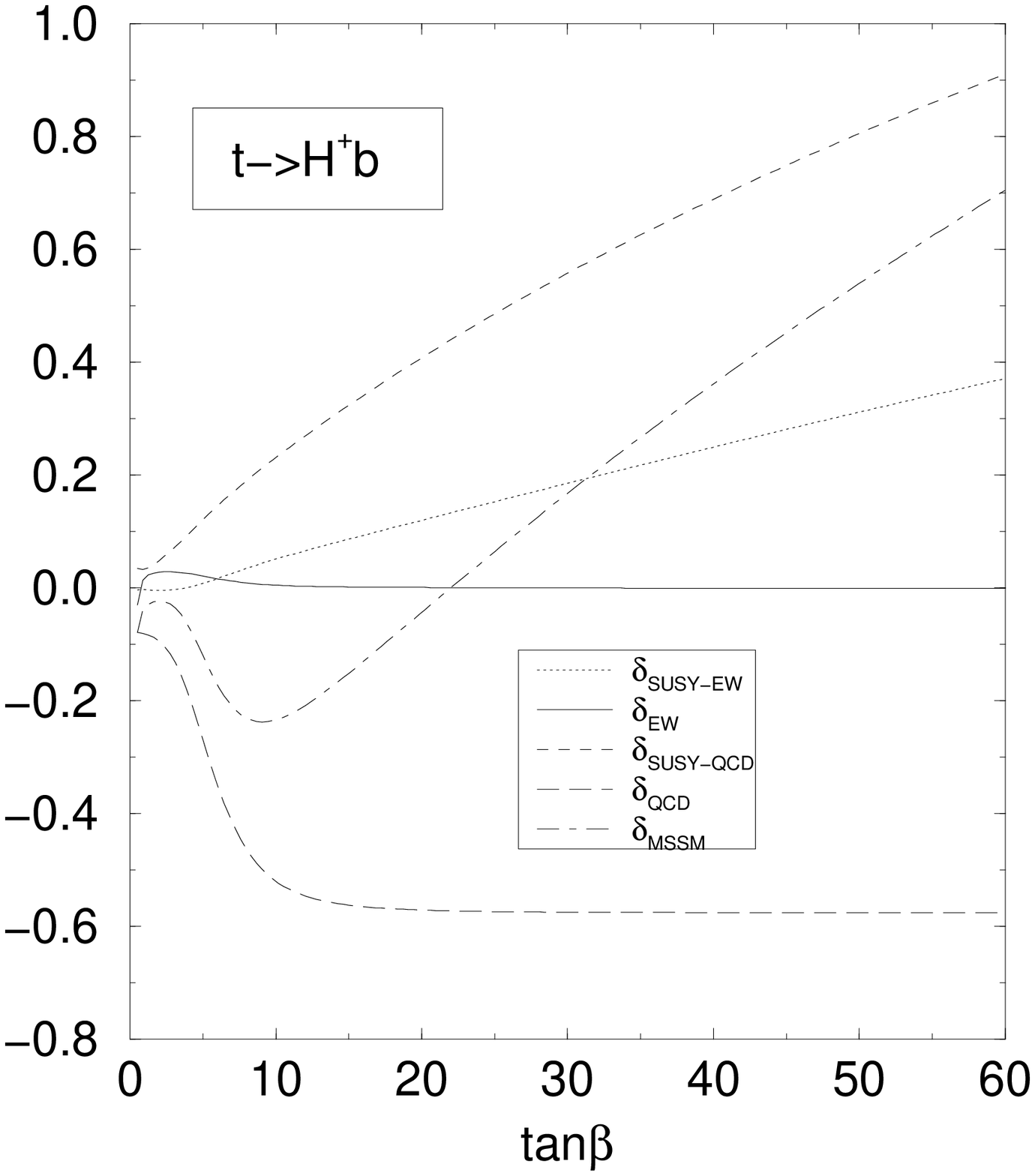,width=9cm}} &
\mbox{\epsfig{file=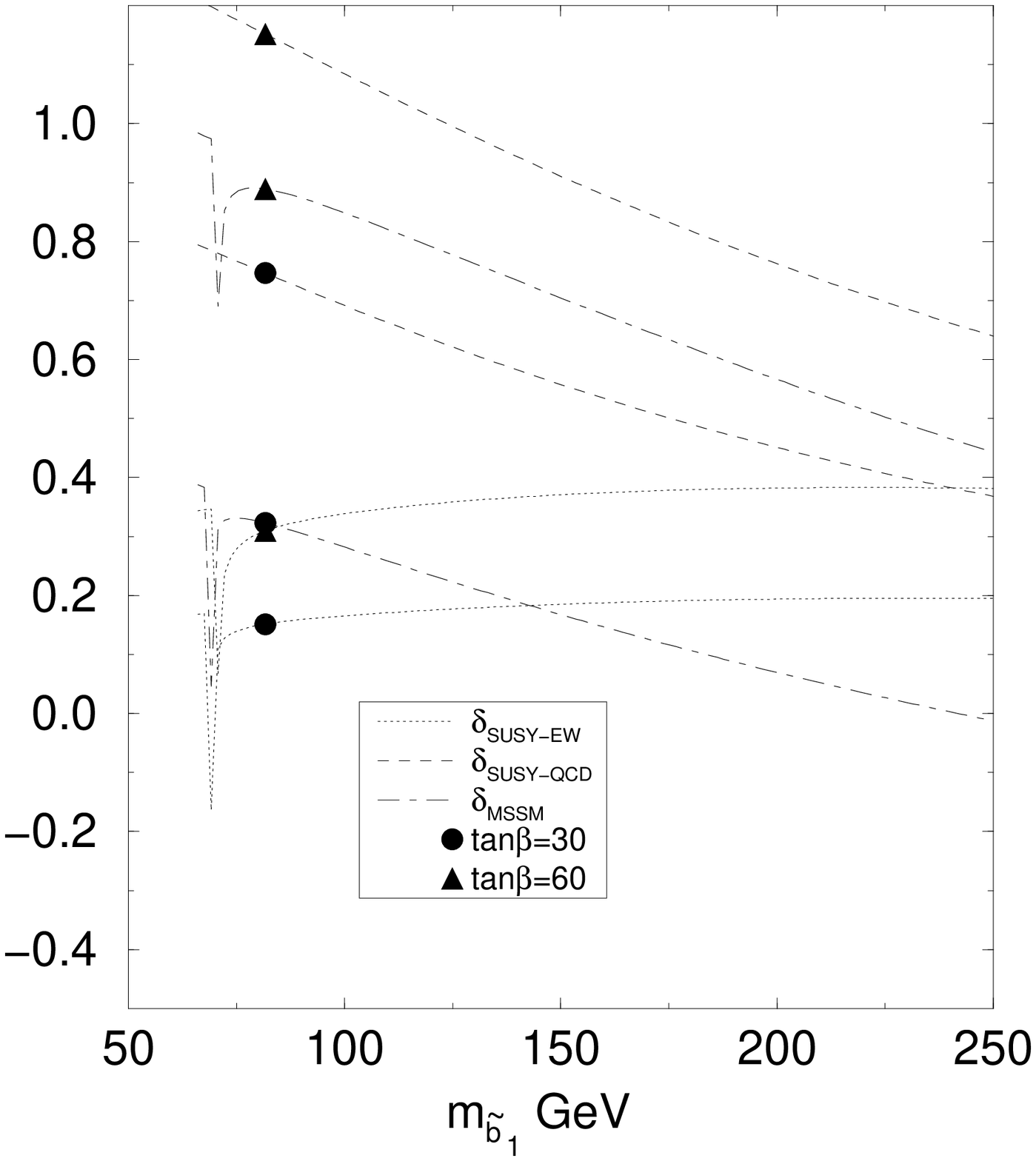,width=9cm}} \\
(a) & (b) \\
~&~\\
\mbox{\epsfig{file=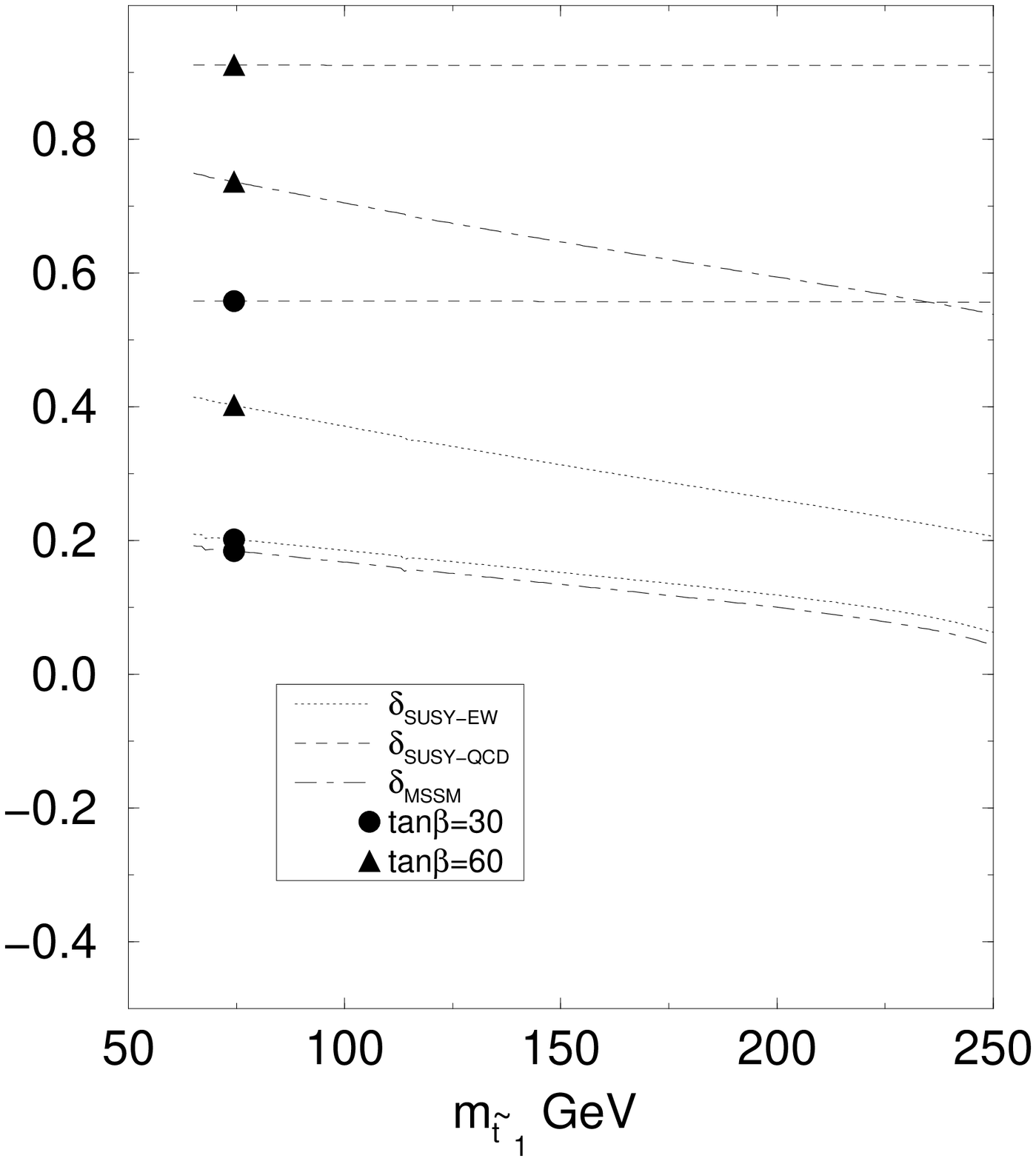,width=9cm}} &
\mbox{\epsfig{file=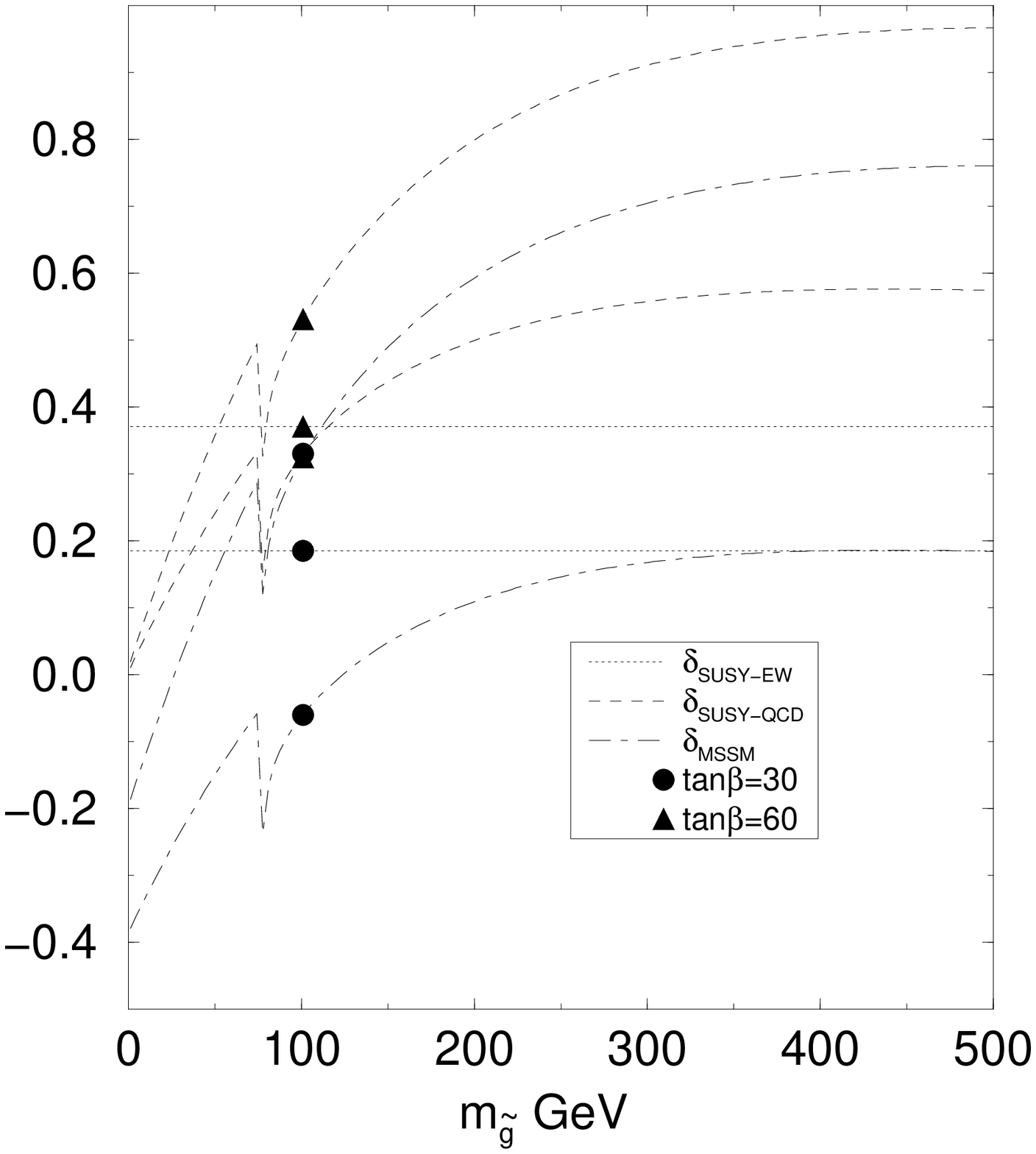,width=9cm}} \\
(c) & (d) 
\end{tabular}}
\begin{center}
{\Huge Fig.3}
\end{center}
\newpage

\centerline{
\begin{tabular}{c}
\mbox{\epsfig{file=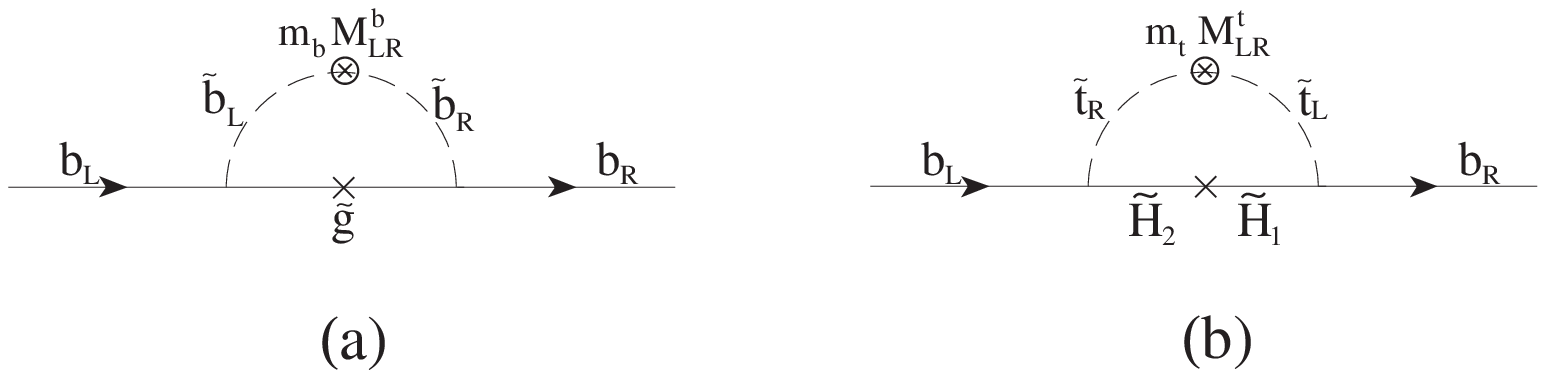,width=18cm}}\\
~\\
\mbox{\epsfig{file=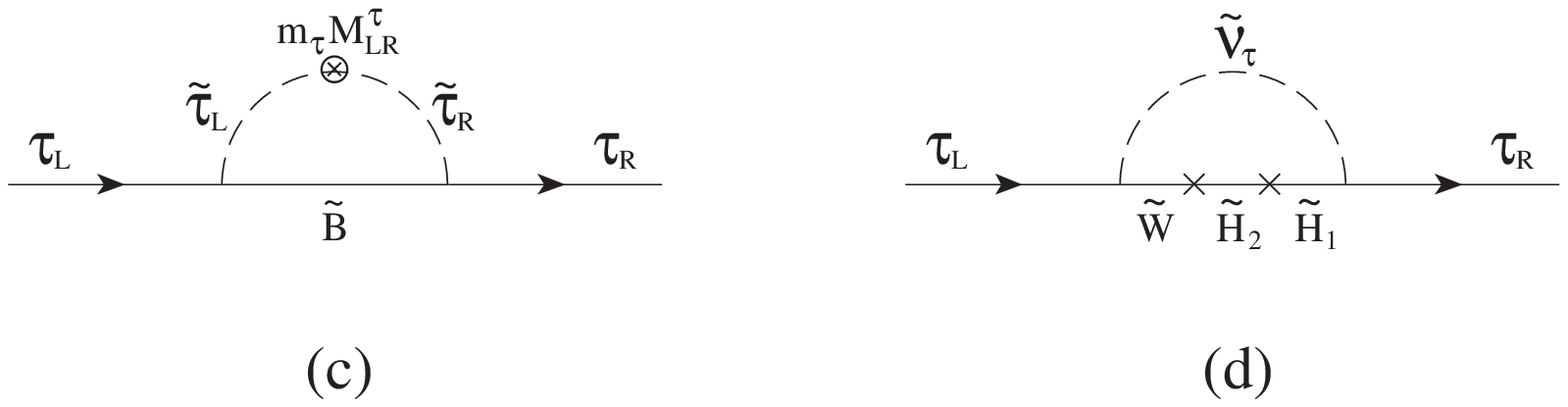,width=18cm}}
\end{tabular}}
\begin{center}
{\Huge Fig.4}
\end{center}

\centerline{
\begin{tabular}{c@{\hspace{1cm}}c}
\mbox{\epsfig{file=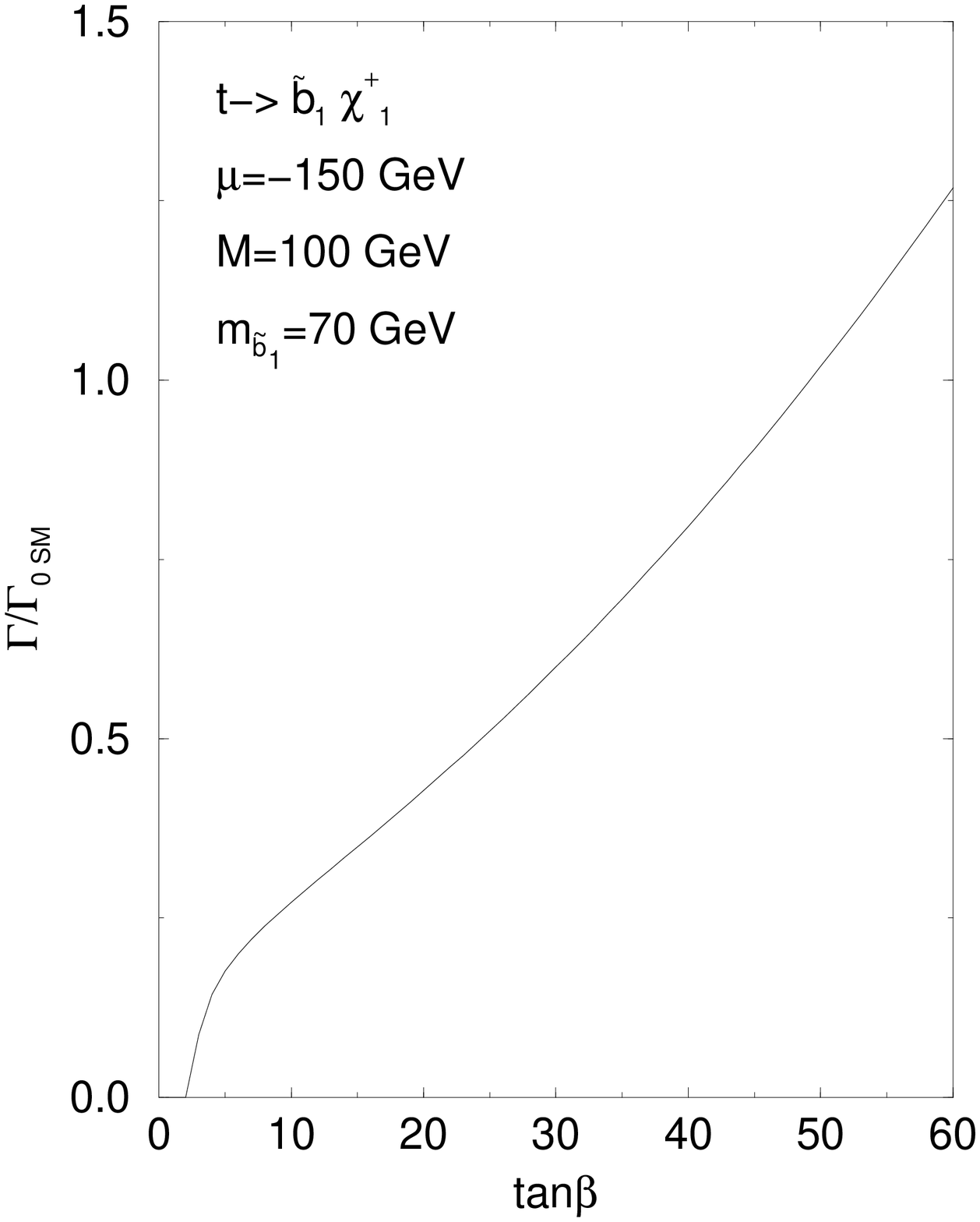,width=9cm}} &
\mbox{\epsfig{file=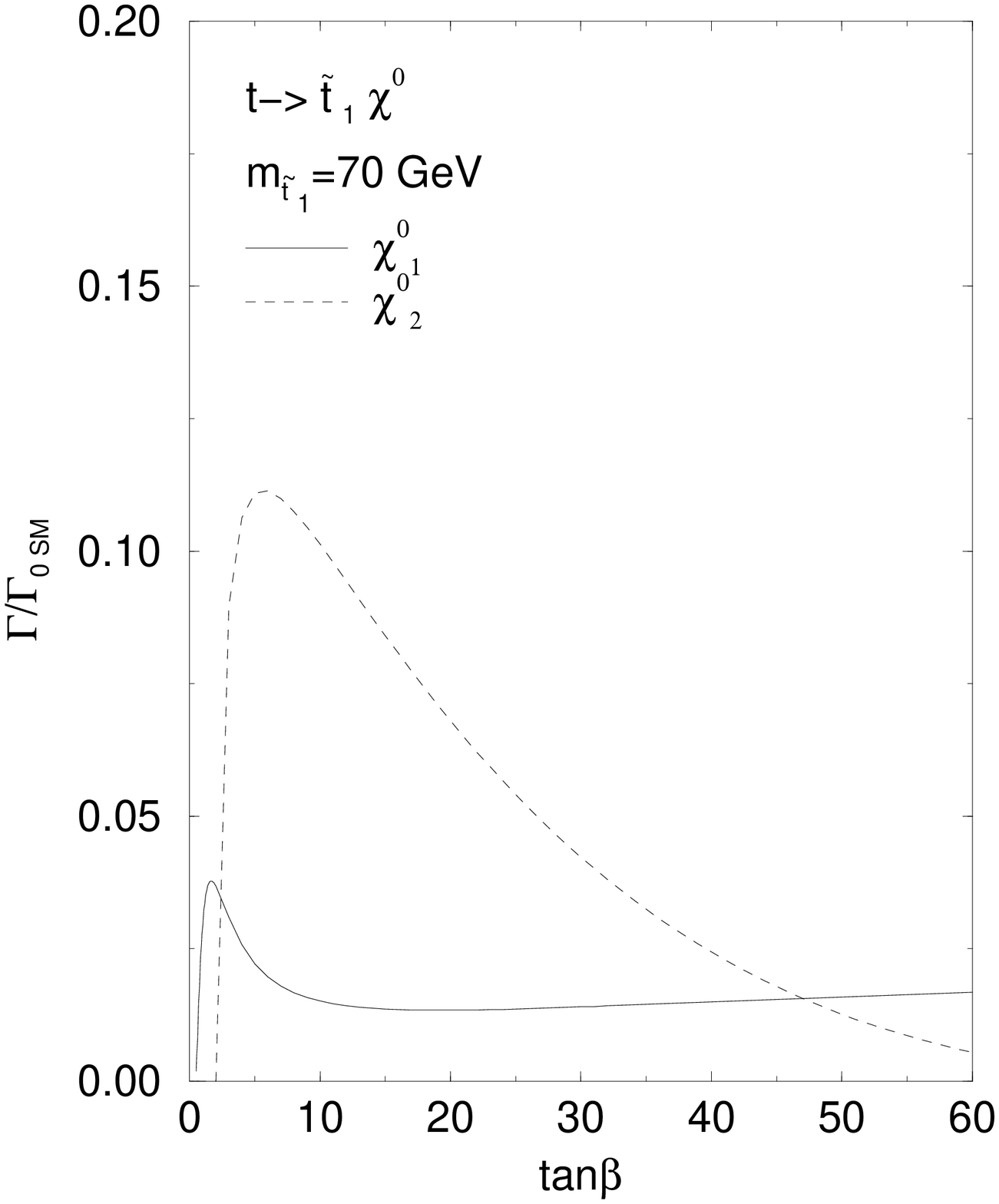,width=9cm}}\\
(a) & (b) \\
\mbox{\epsfig{file=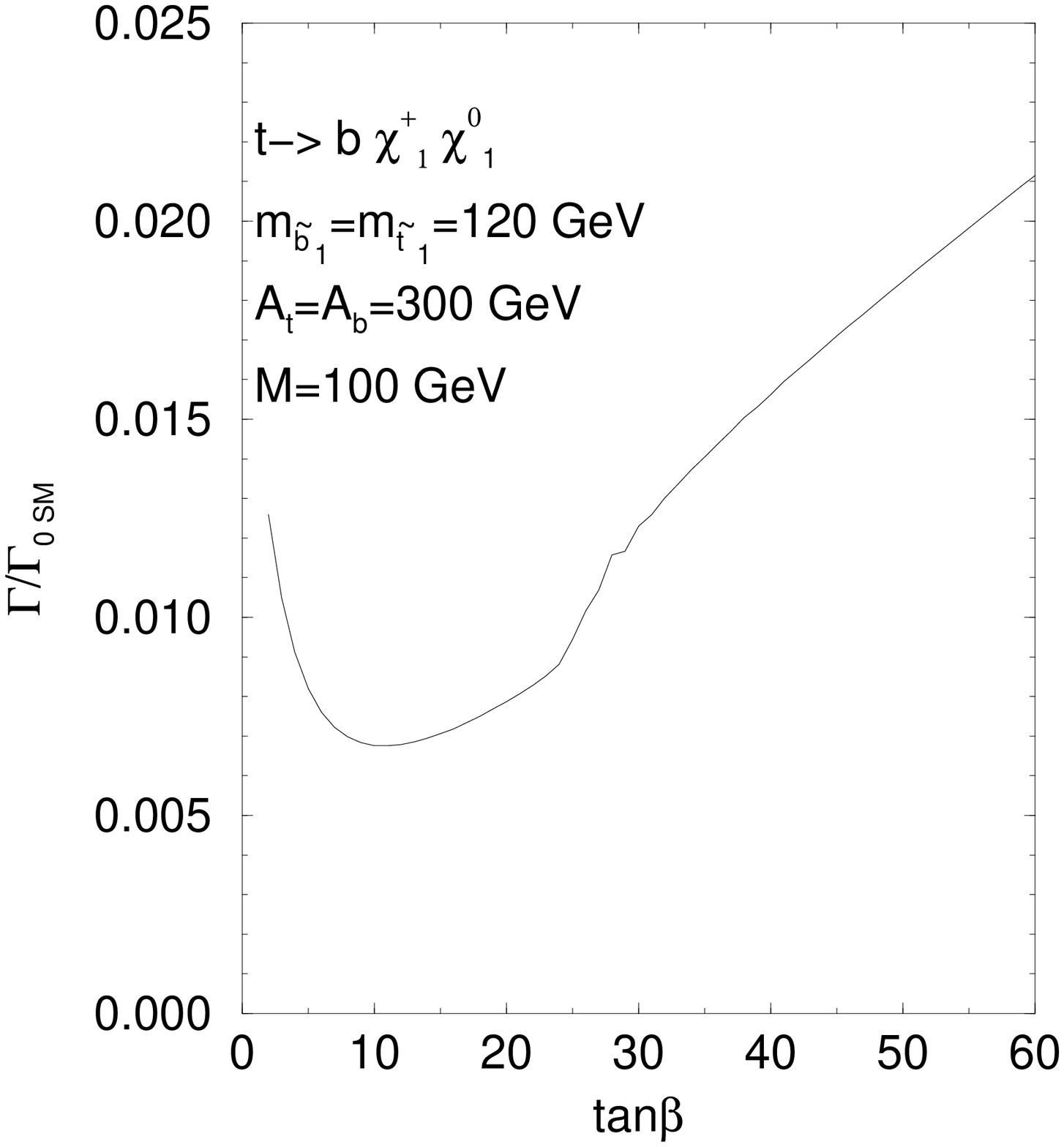,width=9cm}} &
\mbox{\epsfig{file=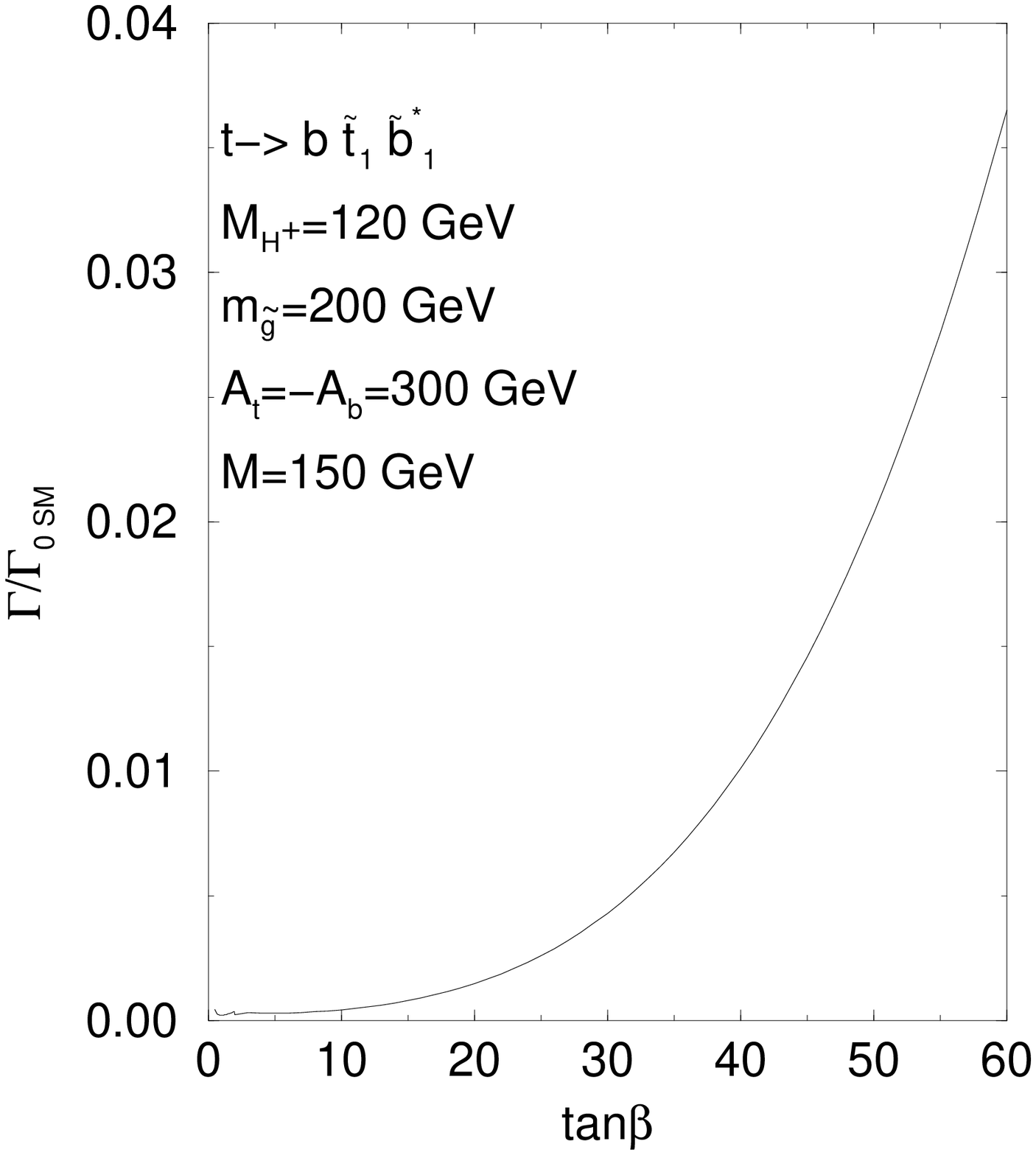,width=9cm}} \\
(c) & (d) 
\end{tabular}}
\begin{center}
{\Huge Fig.5}
\end{center}

\begin{tabular}{c}
\epsfig{file=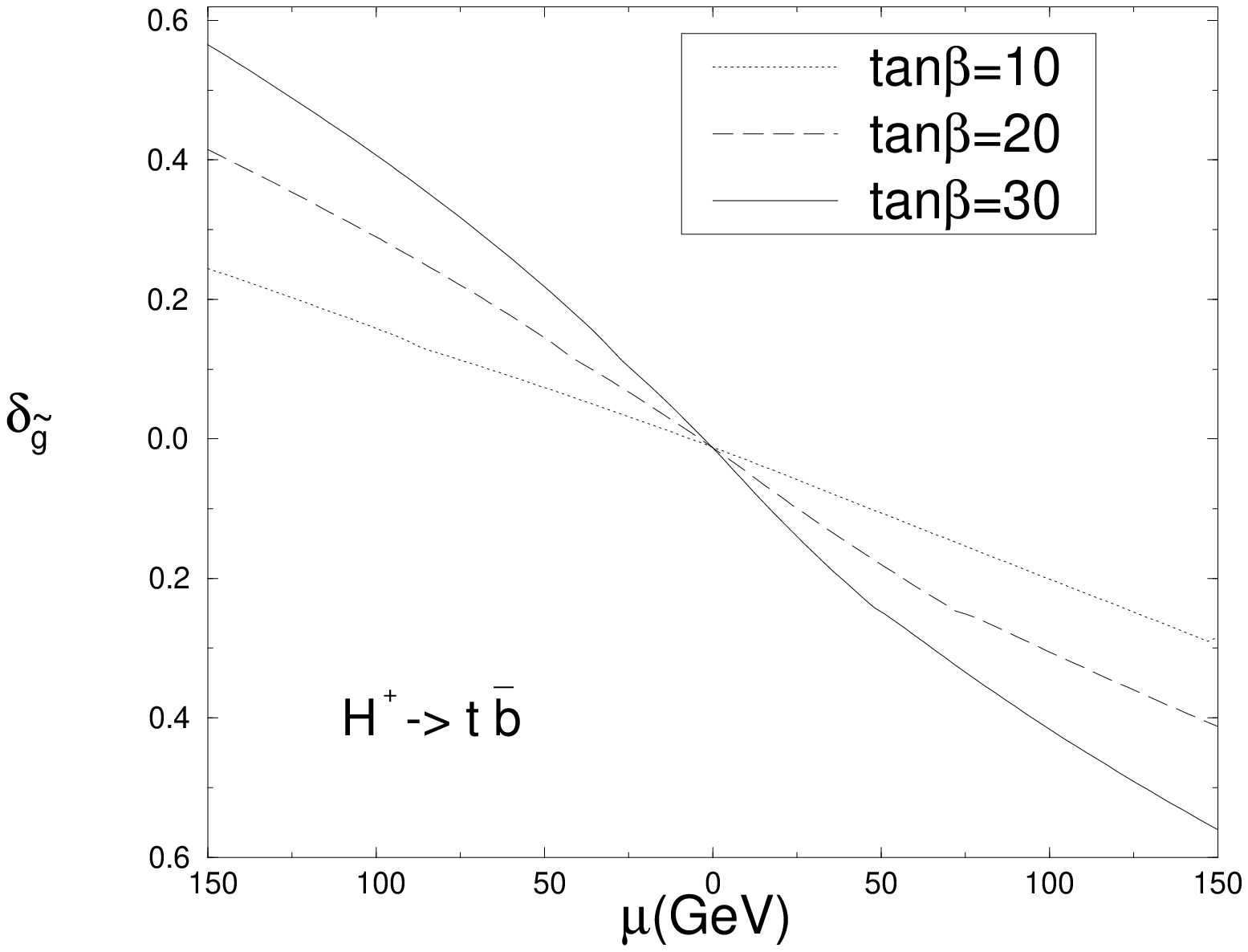,height=11cm}  \\
(a)\\
\epsfig{file=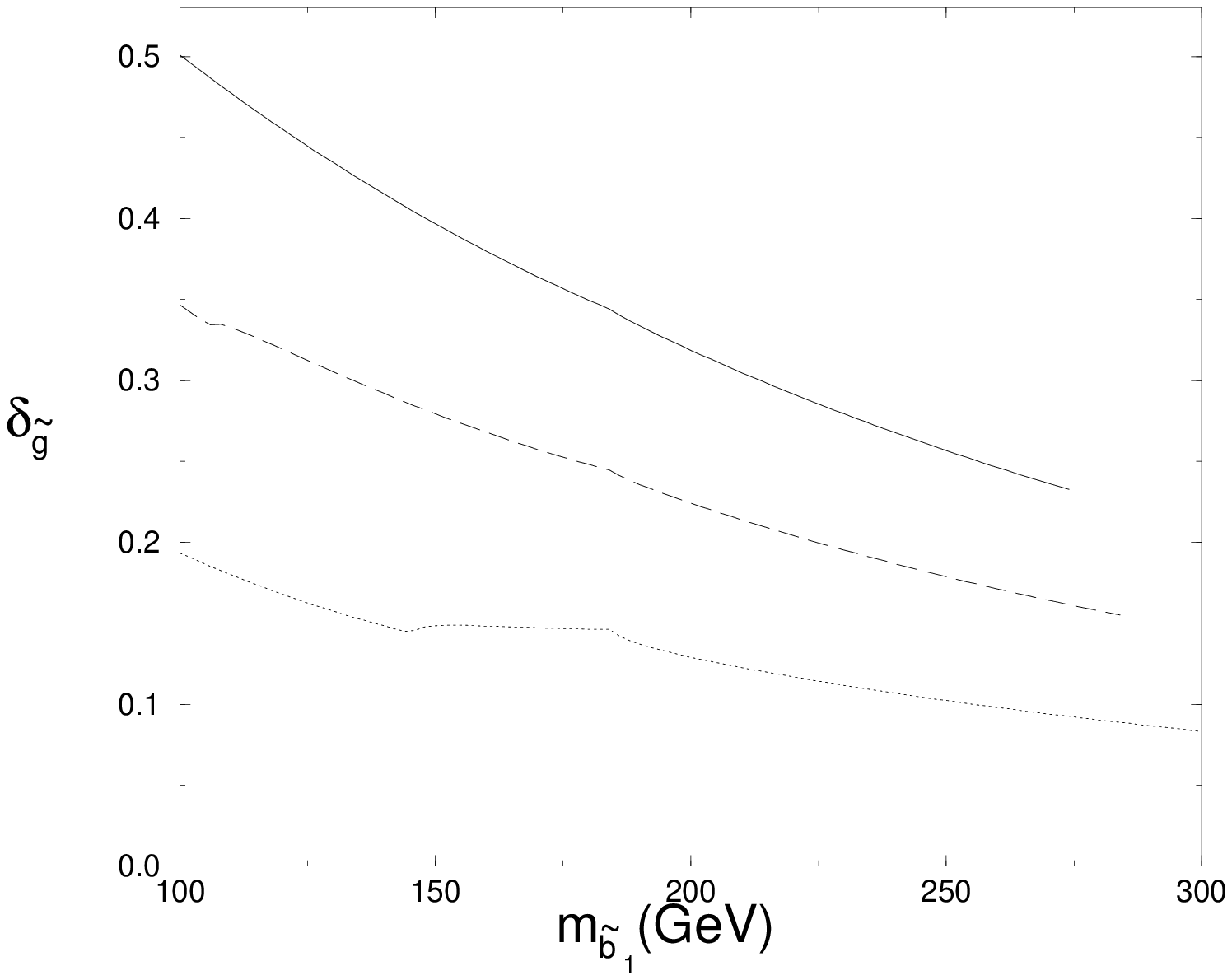,height=11cm}\\
(b)\\
\end{tabular}
\begin{center}
  {\Huge Fig.6}
\end{center}

\begin{tabular}{c}
\epsfig{file=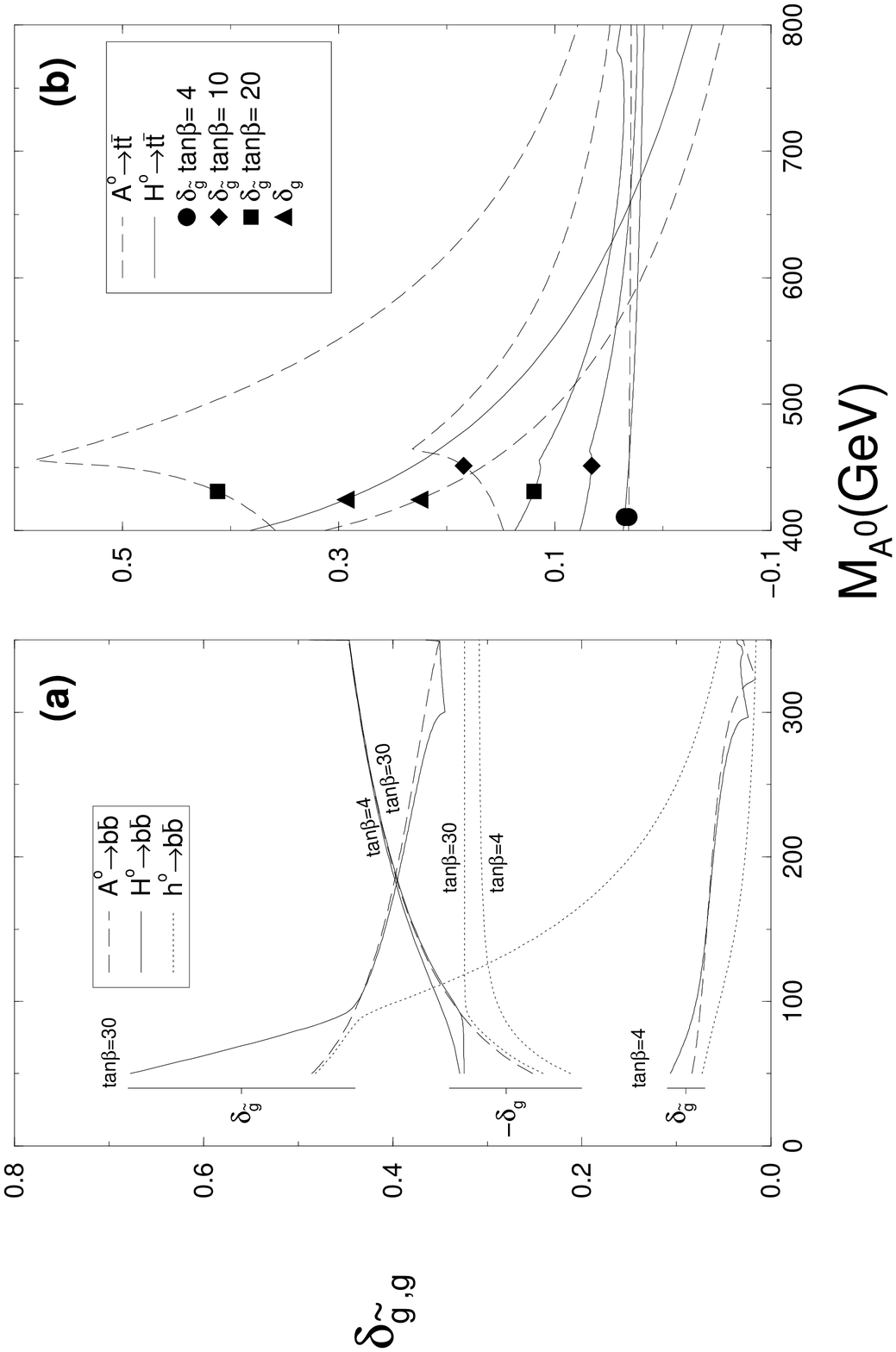,width=14cm}
\\
~\\
~\\
{\Huge Fig.7}
\end{tabular}

\centerline{
\mbox{\epsfig{file=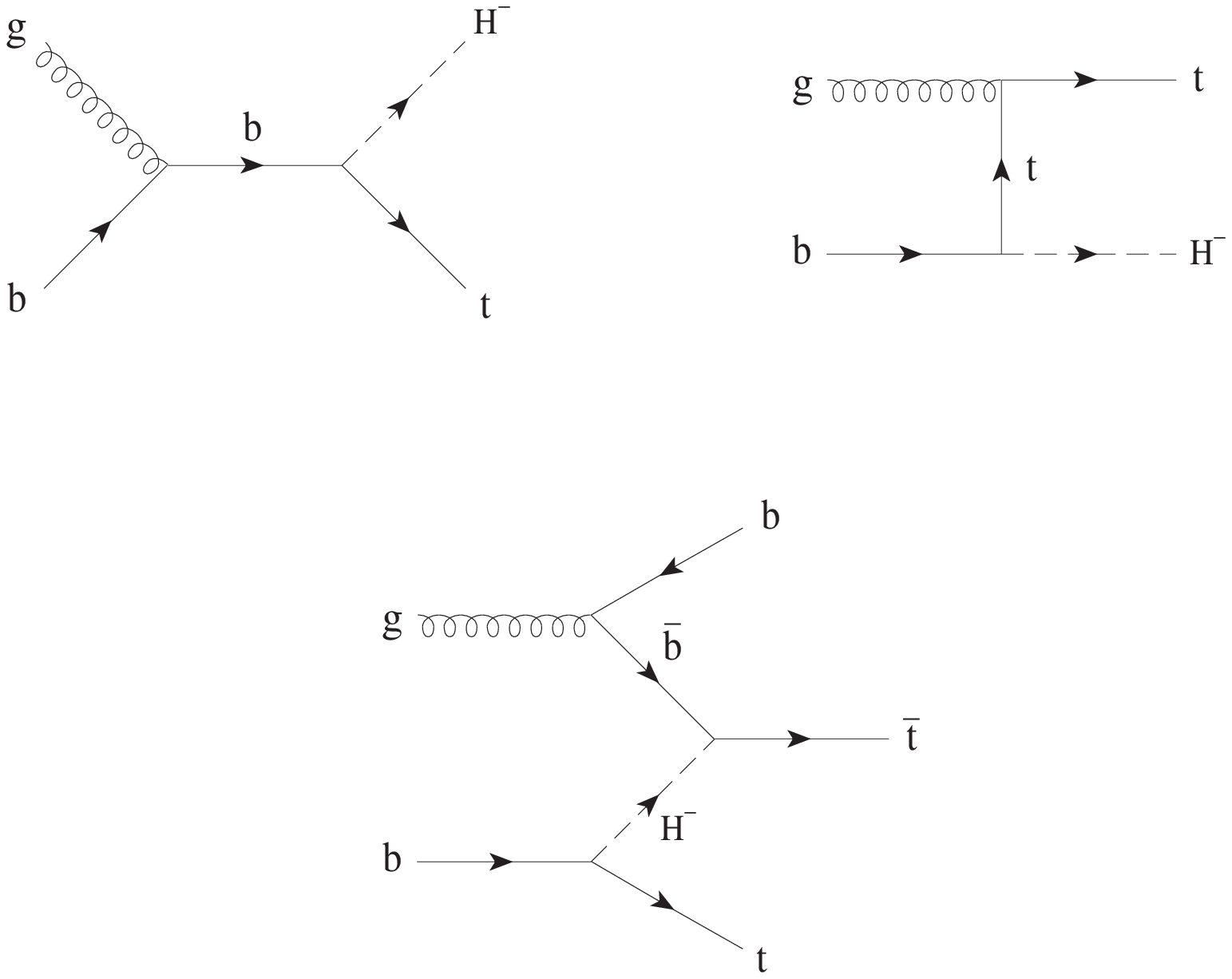,width=18cm}} }
\begin{center}
{\Huge Fig.8}
\end{center}

\end{document}